\documentstyle[12pt]{article}

\catcode `\@=11 \@addtoreset{equation}{section}
\def\theequation{\arabic{section}.\arabic{equation}}
\catcode `\@=12

\newcommand{\be}{\begin{equation}}
\newcommand{\en}{\end{equation}}
\newcommand{\bea}{\begin{eqnarray}}
\newcommand{\ena}{\end{eqnarray}}
\newcommand{\beano}{\begin{eqnarray*}}
\newcommand{\enano}{\end{eqnarray*}}
\newcommand{\bee}{\begin{enumerate}}
\newcommand{\ene}{\end{enumerate}}

\newcommand{\Z}{Z \!\!\!\!\! Z}

\newcommand{\Id}{1\!\!1}

\newcommand{\E}{{\cal E}}

\newcommand{\1}{1 \!\!\! 1}

\begin{document}

\thispagestyle{empty}

\vspace*{1cm}

\begin{center}
{\Large \bf Many-body applications of the stochastic
limit: a review}   \vspace{2cm}\\

{\large F. Bagarello}
\vspace{3mm}\\
  Dipartimento di Metodi e Modelli Matematici,
Facolt\`a di Ingegneria, Universit\`a di Palermo, \\Viale delle Scienze, I-90128  Palermo, Italy\\
e-mail: bagarell@unipa.it\\home page: www.unipa.it/$^\sim$bagarell
\vspace{4mm}\\

\end{center}

\vspace*{2cm}

\begin{abstract}
\noindent We review some applications of the perturbative
technique known as the {\em stochastic limit approach} to the
analysis of the following many-body problems: the fractional
quantum Hall effect, the relations between the Hepp-Lieb and the
Alli-Sewell models (as possible models of interaction between
matter and radiation), and the open BCS model of low temperature
superconductivity.
\end{abstract}

\vspace{2cm}

\noindent
{\bf Keywords}: Stochastic limit. BCS model. Quantum Hall effect. Laser models. \\
{\bf PACS Numbers}: 02.90.+p, 03.65.Db \vfill

\newpage
\section{Introduction}

In this paper we review with a certain care the main results
concerning three applications of the so-called {\em stochastic
limit approach} (SLA), see \cite{book} for the main definitions
and some physical applications and \cite{acmat} for some rigorous
mathematical results, to three problems in quantum mechanics for
systems with infinite degrees of freedom. In particular we devote
Section II to the discussion of the fractional quantum Hall
effect, \cite{bagacc}. In Section III we discuss many relations
between different models of matter interacting with radiation: the
Hepp-Lieb and the Dicke-Haken-Lax hamiltonian models, and the
Alli-Sewell dissipative counterpart, \cite{baglaser}. In Section
IV, we discuss the open BCS model of superconductivity as
originally proposed by Martin and Buffet, \cite{bm}, and some
quite recent results related to that, \cite{bagbcs,bagbcs2}. For
reader's convenience, we also add a brief Appendix where some
crucial facts on the SLA are listed.

\section{The fractional quantum Hall effect}

The Hamiltonian for the quantum Hall effect (QHE) is, see for
instance reference \cite{BMS},
\begin{equation}
H^{(N)}=H^{(N)}_0+\lambda(H^{(N)}_c+H^{(N)}_B)\label{6.1}
\end{equation}
where $H^{(N)}_0$ is the Hamiltonian for the free $N$ electrons,
$H^{(N)}_c$ is the  Coulomb interaction:
\begin{equation}
H^{(N)}_c={1\over2}\,\sum^N_{i\not=j}{e^2\over|\underline r_i-
\underline r_j|}\label{6.4}
\end{equation}
and $H^{(N)}_B$ is the interaction of the charges with the
positive uniform background. A huge literature exists concerning
the QHE. We suggest here references \cite{cha} and \cite{gir}.

In this section we consider a model defined by an Hamiltonian
\begin{equation}
H=H_0^{(N)}+H_{0,R}+\lambda H_{eb}=H_0+\lambda H_{eb}\label{ham}
\end{equation}
which is obtained from the Hamiltonian (\ref{6.1}) by introducing
the following approximations:

$(1)$ the Coulomb background-background interaction is replaced by
the free bosons Hamiltonian $H_{0,R}$ given by
\begin{equation}
H_{0,R}=\int\omega(\underline k)b^+(\underline k)b(\underline k)d
k\label{37h0}
\end{equation}
where $\omega(\underline k)$ is the dispersion  for the free
background and $b^+(\underline k)$ and $b(\underline k)$ are
bosonic operators satisfying the canonical comutation relations:
\begin{equation}
[b(\underline k),b^+(\underline k')]=\delta(\underline
k-\underline k')\quad[b(\underline k),b(\underline
k')]=[b^+(\underline k),b^+(\underline k')]=0 \label{ccr}
\end{equation}

$(2)$ the Coulomb electron-electron and electron-background
interaction is replaced by the Fr\"ohlich Hamiltonian  $H_{eb}$,
\cite{Str},  which is only quadratic rather than quartic in the
fermionic operators:

\begin{equation}
H_{eb}=\int\psi^\dagger(\underline r)\psi(\underline r)\tilde
F(\underline r-\underline r')\phi(\underline r')d rd r'\label{33}
\end{equation}
where $\psi(\underline r)$ and $\phi(\underline r')$ are
respectively the electron and the bosonic fields, while $\tilde F$
is a form factor. Expanding $\phi (\underline r)$ in plane waves,
$\psi(\underline r)$ in terms of the eigenstates
$\psi_\alpha(\underline r)$ of the single electron hamiltonian,
see (\ref{15}) below, introducing the form factors
\begin{equation}
g_{\alpha\beta}(\underline k):={1\over\sqrt{(2\pi)^3}}\,{\hat
V_{\alpha \beta}(\underline k)\over\sqrt{2\omega(\underline
k)}}\label{34}
\end{equation}
where
\begin{equation}
\hat V_{\alpha\beta'}(\underline
k):=\int\overline{\psi_\alpha(\underline r)} e^{i\underline
k\cdot\underline r}\psi_{\beta'}(\underline r)d r\label{19}
\end{equation}
and taking $\tilde F(\underline r)=e^2 \delta(\underline r)$,
\cite{Str}, we can write
\begin{equation}
H_{eb}=e^2 \sum_{\alpha\beta}a^+_\alpha
a_\beta(b(g_{\alpha\beta})+b^+
(\overline{g_{\beta\alpha}}))\label{35eb}
\end{equation}
which is quadratic in the fermionic operators $a_\alpha$,
$a^+_\alpha$,
\begin{equation}
\{a_\alpha,a_\beta\}=\{a^+_\alpha,a^+_\beta\}=0\qquad\{a_\alpha,a^+_\beta\}=
\delta_{\alpha\beta}\label{20}
\end{equation}
Here we have introduced the smeared operators
\begin{equation}
b(g_{\beta\alpha})=\int dk\, b(\underline
k)\,g_{\beta\alpha}(\underline k).
\end{equation}

Notice that we are adopting here and in the following the
simplifying notation for the quantum numbers
$\alpha:=(n_\alpha,p_\alpha)$, see formula (\ref{15}) below.

These are certainly strong approximations. However since from the
Fr\"ohlich Hamiltonian it is possible to recover a quartic
interaction,  \cite{Str}, one can say that the Fr\"ohlich
Hamiltonian describes an effective electron-electron interaction
which may mimic at least some aspects of the original Coulomb
interaction. From this point of view it seems natural to
conjecture that some dynamical phenomena deduced from this
Hamiltonian might have implications in the study of the  QHE. This
conjecture is supported by our main result, given by formulae
(\ref{120}) and (\ref{121}) where we deduce, directly from the
dynamics, an obstruction to the presence of a non zero
$x$-component of the current, which is quantized according to the
values of a finite set of rational numbers.  This is what we will
call a {\it fine tuning condition} (FTC).

Useless to say that the FTC  strongly reminds the rational values
of the filling factor  for which the plateaux are observed in the
real QHE. We will comment again this fact later.

In these notes we discuss a model of $N<\infty$ charged
interacting particles concentrated around a two dimensional layer
contained in the $(x,y)$-plane and subjected to a uniform electric
field $\underline E=E\hat j$, along $y$, and to an uniform
magnetic field $\underline B=B\hat k$ along $z$.

The Hamiltonian for the free $N$ electrons $H^{(N)}_0$, is the sum
of $N$ contributions:
\begin{equation}
H^{(N)}_0=\sum^N_{i=1}H_0(i)\label{2}
\end{equation}
where $H_0(i)$ describes the minimal coupling of the $i$--th
electrons with the field:
\begin{equation}
H_0(i)={1\over2m}\,\left(\underline p+{e\over c}\,\underline
A(r_i)\right)^2+e \underline E\cdot\underline r_i\label{3}
\end{equation}

In the Landau gauge $\underline A=-B(y,0,0)$, and requiring
periodic boundary condition on $x$,
$\psi(-L_x/2,y)=\psi(L_x/2,y)$, for almost all $y$, we find
\begin{equation} \psi_{np}(\underline r)={e^{i{2\pi px\over
L_x}}\over\sqrt{L_x}}\,
\varphi_n(y-y^{(p)}_0)\quad\quad\varepsilon_{np}=\hbar\omega(n+1/2)-{eE
\over2m\omega^2}\,\left(eE-{4\hbar\omega\pi p\over
L_x}\right)\label{15}
\end{equation}
where $\varphi_n$ is the $n$--th eigenstate of the one-dimensional
harmonic oscillator, $\omega=\frac{eB}{mc}$,
$y_0=\frac{1}{m\omega^2}(\hbar k\omega-eE)$ and $k={2\pi\over
L_x}\,p$, where $p\in{\bf Z}$.

Equation (\ref{15}) shows that the wave function
$\psi_{np}(\underline r)$ factorizes in a $x$--dependent part,
which is labelled by the quantum number $p$, and a part,  only
depending on $y$, which is labelled by both $n$ {\bf and\/} $p$
due to the presence of $y^{(p)}_0$ in the argument of the function
$\varphi_n$.

Of course the  Hamiltonian $H^{(N)}_0$ in (\ref{2}) becames, in a
second quantized form,
\begin{equation}
H_0^{(N)}=\sum_\alpha\varepsilon_\alpha a^+_\alpha
a_\alpha,\label{36h0}
\end{equation}
where the $\varepsilon_\alpha$ are the single electron energies,
labeled by the pairs $\alpha=(n,p)$ as given in formula (\ref{15})
and the $a_\alpha^\sharp$ are the fermionic operators satisfying
(\ref{20}).

\subsection{The stochastic limit of the model}

In this subsection we briefly outline how to apply the stochastic
limit procedure to the model introduced above. The stochastic
limit describes the dominating contribution to the dynamics in
time scales of the order $t/\lambda^2$, where $\lambda$ is the
coupling constant.

The starting point is the Hamiltonian (\ref{ham}) together with
the commutation relations (\ref{20}), (\ref{ccr}). Of course, the
Fermi and the Bose operators commute among them. The interaction
Hamiltonian $H_{eb}$ for this model is given by (\ref{35eb}) and
the free Hamiltonian $H_0$ is given by (\ref{ham}), (\ref{37h0})
and (\ref{36h0}).

The time evolution of $H_{eb}$, in the interaction picture is then
\begin{equation}
H_{eb}(t)=e^{iH_0t}H_{eb}e^{-iH_0t}=e^2\sum_{\alpha\beta}a^+_\alpha
a_\beta(b(g_{\alpha\beta}e^{-it(\omega-\varepsilon_{\alpha\beta})})+
b^+(\overline
g_{\beta\alpha}e^{it(\omega-\varepsilon_{\beta\alpha})}))\label{75}
\end{equation}
where
\begin{equation}\varepsilon_{\alpha\beta}=\varepsilon_\alpha-\varepsilon_\beta\label{43}
\end{equation}

Therefore the Schr\"odinger equation in the interaction
representation is:
\begin{equation}
\partial_tU^{(\lambda)}_t=-i\lambda H_{eb}(t)U^{(\lambda)}_t,\label{47}
\end{equation}
which becames, after the time rescaling $t\to t/\lambda^2$,
\begin{equation}
\partial_tU^{(\lambda)}_{t/\lambda^2}=-{i\over\lambda} H_{eb}(t/\lambda^2)
U^{(\lambda)}_{t/\lambda^2}.\label{48}
\end{equation}
Its integral form is
\begin{equation}
U^{(\lambda)}_{t/\lambda^2}=1\!\!\!\!1-{i\over\lambda}\,\int^t_0H_{eb}
(t'/\lambda^2)U^{(\lambda)}_{t'/\lambda^2}dt'\label{49}
\end{equation}

We see that the rescaled Hamiltonian
\begin{equation}
{1\over\lambda}\,H_{eb}(t/\lambda^2)=e^2\,\sum_{\alpha \beta}
a_\alpha^\dagger a_\beta{1\over
\lambda}\,b\left(e^{-it\over\lambda^2}\,(\omega-\varepsilon_{\alpha
\beta}) g_{\alpha \beta}\right)+\hbox{ h.c.}\label{50}
\end{equation}
depends on the rescaled fields
\begin{equation}
b_{\alpha
\beta,\lambda}(t)={1\over\lambda}\,b(e^{-i{t\over\lambda^2}\,
(\omega- \varepsilon_{\alpha \beta})}g_{\alpha \beta})\label{51}
\end{equation}
The first statement of the stochastic golden rule, \cite{book}, is
that the rescaled fields converge (in the sense of correlators) to
a quantum white noise
\begin{equation}
b_{\alpha\beta}(t)=\lim_{\lambda\to0}{1\over\lambda}\,b(g_{\alpha
\beta} e^{-i{t\over\lambda^2} \,(\omega-\varepsilon_{\alpha
\beta})})\label{52}
\end{equation}
characterized by the following commutation relations
\begin{equation}
[b_{\alpha\beta}(t),b_{\alpha'\beta'}(t')]=[b^+_{\alpha\beta}(t),
b^+_{\alpha'\beta'}(t')]=0 \label{82zz}
\end{equation}
\begin{equation}
[b_{\alpha\beta}(t),b^+_{\alpha'\beta'}(t')]=
\delta_{\varepsilon_{\alpha\beta},\varepsilon_{\alpha'\beta'}}\delta(t-t')
G^{\alpha\beta\alpha'\beta'} \label{82zzz}
\end{equation}
where the constants $G^{\alpha\beta\alpha'\beta'}$ are given by
\begin{equation}
G^{\alpha\beta\alpha'\beta'}=\int^\infty_{-\infty}d\tau\int d
kg_{\alpha \beta}(\underline
k)\overline{g_{\alpha'\beta'}(\underline
k)}e^{i\tau(\omega(\underline k)-\epsilon_{\alpha\beta})}=2\pi\int
dk g_{\alpha\beta}(k)g_{\alpha\beta}(k)\delta(\omega(\underline
k)-\varepsilon_{\alpha \beta})\label{78}
\end{equation}
The vacuum of the master fields $b_{\alpha\beta}(t)$ will be
denoted by $\eta_0$:
\begin{equation}
b_{\alpha\beta}(t)\eta_0=0\quad\forall\,\alpha \, \beta,\
\forall\,t\label{54}
\end{equation}
The limit Hamiltonian is, then, see \cite{bagacc},
\begin{equation}
H^{(sl)}_{eb}(t)=e^2\sum_{\alpha\beta}(a^+_\alpha a_\beta
b_{\alpha \beta}(t)+\hbox{ h.c.})\label{76}
\end{equation}
In this sense we say that $H^{(sl)}_{eb}(t)$ is the ``stochastic
limit'' of $H_{eb}(t)$ in (\ref{75}). Moreover, the wave operator
in the stochastic limit satisfies the equation
\begin{equation}
\partial_tU_t=-iH^{(sl)}_{eb}(t)U_t\label{56}
\end{equation}
or, in integral form,
\begin{equation}
U_t=\Id-i\int^t_0H^{(sl)}_{eb}(t')U_{t'}dt',\label{57}
\end{equation}

Finally, the stochastic limit of the (Heisenberg) time evolution
of any observable $X$ of the system is:
\begin{equation}
j_t(\tilde X)=U^+_t(X\otimes\Id_R)U_t,\label{58}
\end{equation}
where $\Id_R$ is the identity of the reservoir. Since the
$b_{\alpha\beta}(t)$ are quantum white noises, equation
(\ref{56}), and the corresponding differential equation for
$j_t(\tilde X)$, are singular equations and to give them a meaning
we bring them in normal form. This normally ordered evolution
equation is called {\it the quantum Langevin equation\/}. Its
explicit form is:
$$\partial_tj_t( X)=e^2\sum_{\alpha\beta}\{j_t([a^+_\alpha
a_\beta, X]\Gamma^{\alpha\beta}_--\Gamma^{\alpha\beta}_-
[a^+_\beta a_\alpha, X])\}+$$
\begin{equation}
+ie^2\sum_{\alpha\beta}\{b^+_{\alpha\beta}(t)j_t([a^+_\beta
a_\alpha, X]) +j_t([a^+_\alpha a_\beta,
X])b_{\alpha\beta}(t)\}\label{80}
\end{equation}
where
\begin{equation}
\Gamma^{\alpha\beta}_-:=\sum_{\alpha'\beta'}\delta_{\varepsilon_{\alpha
\beta},\varepsilon_{\alpha'\beta'}}a^+_{\beta'}a_{\alpha'}
G^{\alpha\beta\alpha'\beta'}_-\label{81}
\end{equation}
\begin{equation}
G_-^{\alpha\beta\alpha'\beta'}=\int^0_{-\infty}d\tau\int d
kg_{\alpha\beta} (\underline
k)\overline{g_{\alpha'\beta'}(\underline
k)}e^{i\tau(\omega(\underline
k)-\epsilon_{\alpha\beta})}=\label{77}
\end{equation}
$$={1\over2}\,G^{\alpha\beta\alpha'\beta'}-i\hbox{ P.P. }\int
g^{(k)}_{\alpha\beta}\overline{g^{(k)}_{\alpha'\beta'}}{1\over
\omega_k-\varepsilon_{\alpha\beta}}$$

The master equation is obtained by taking the mean value of
(\ref{80}) in the state $\eta_0^{(\xi)}=\eta_0\otimes\xi$, $\xi$
being a generic vector of the system. This gives
\begin{equation}
\langle\partial_t j_t( X)\rangle_{\eta^{(\xi)}_0}=e^2\sum_{\alpha
\beta}\langle j_t([a^+_\alpha a_\beta, X]\Gamma^{\alpha\beta}_-
-\Gamma^{\alpha\beta+}_-[a^+_\beta a_\alpha,
X])\rangle_{\eta^{(\xi)}_0}\label{82}
\end{equation}
and from this we find for the generator
\begin{equation}
L(
X)=e\sum_{\alpha\beta\alpha'\beta'}\delta_{\varepsilon_{\alpha\beta},
\varepsilon_{\alpha'\beta'}}\{[a^+_\alpha a_\beta, X]a^+_{\beta'}
a_{\alpha'}G^{\alpha\beta\alpha'\beta'}_-
-a^+_{\alpha'}a_{\beta'}[a^+_\beta a_\alpha, X]
\overline{G^{\alpha\beta\alpha'\beta'}_-}\}\label{83}
\end{equation}

This expressions for $L( X)$  will be the starting point for our
successive analysis.

\subsection{The current operator in second quantization}

The current is proportional to the sum of the velocities of the
electrons:
\begin{equation}
\vec J_\Lambda(t)=\alpha_c\sum^{N}_{i=1}{d\over dt}\,\vec
R_i(t).\label{84}
\end{equation}
Here $\Lambda$ is the two--dimensional region corresponding to the
physical layer, $\alpha_c$ is a proportionality constant which
takes into account the electron charge, the area of the surface of
the physical device and other physical quantities, and $\vec
R_i(t)$ is the position operator for the $i$--th  electron.
Moreover $N$ is the number of electrons contained in $\Lambda$.
Defining
\begin{equation}
\vec X_\Lambda(t)=\sum_{i=1}^{N}\vec R_i(t)\ ,\label{85}
\end{equation}
we simply write
\begin{equation}
\vec J_\Lambda(t)=\alpha_c\dot{\vec X}_\Lambda(t)\ .\label{86}
\end{equation}
Since $\vec X_\Lambda(t)$ is a sum of single-electron operators
its expression in second quantization is given by
\begin{equation}
\vec X_\Lambda=\sum_{\gamma\mu}\vec X_{\gamma\mu}a^+_\gamma
a_\mu\label{87}
\end{equation}
where
\begin{equation}
\vec X_{\gamma\mu}=\langle\psi_\gamma,\vec
X_\Lambda\psi_\mu\rangle=\int \psi_\gamma(\underline r)\underline
r\psi_\mu(\underline r)d r\label{88}
\end{equation}
Recall that the $\psi_\gamma(\underline r)$ are the single
electron wave functions given in (\ref{15}) and that $a_\alpha$
and $a^+_\alpha$ satisfy the anticommutation relations (\ref{20}).
In the LLL, $n_\gamma=n_\mu=0$, we find that, \cite{bagacc},

\begin{equation}
X^{(1)}_{\gamma\mu}=(1-\delta_{p_\mu
p_\gamma})(-1)^{p_\mu-p_\gamma} L_x {e^{-y^2_{p_\mu
p_\gamma}}\over2 \pi i(p_\mu-p_\gamma)}\label{94a}
\end{equation}
\begin{equation}
X^{(2)}_{\gamma\mu}=y^{(p_\gamma)}_0\delta_{p_\mu
p_\gamma}\label{94b}
\end{equation}
where $y_{p_\mu p_\gamma}:=\sqrt{m\omega\over
4\hbar}(y^{(p_\mu)}_0-y^{(p_\gamma)}_0)= {\pi\over
L_x}\,\sqrt{\hbar\over m\omega}(p_\mu-p_\gamma)$.

 To show how these results can be useful in the computation
of the electron current we start noticing that, if $\varrho$ is a
state of the electron system, then
\begin{equation}
\langle\vec J_\Lambda(t)\rangle_\varrho=\alpha_c\langle{d\over
dt}\,\vec X_\Lambda(t)\rangle_\varrho=\alpha_c\langle L(\vec
X_\Lambda(t))\rangle_\varrho=\alpha_c Tr(\varrho{L(\vec
X_\Lambda(t))})\label{95}
\end{equation}

The vector $\langle\vec J_\Lambda(t)\rangle_\varrho$ will be
computed now for a particular class of states $\varrho$, and we
will use this result to get the expressions for the conductivity
and for the resistivity tensor.

To do this we begin computing the electric current. We first need
to find $ L(\vec X_\Lambda)$, $L$ being the generator given in
(\ref{83}). Since $\vec X_\Lambda= \vec X_\Lambda^\dagger$, we
have
$$
L(\vec X_\Lambda)=L_1(\vec X_\Lambda)+h.c.,
$$
where, as we find after a few computations,
\begin{equation}
L_1(\vec X_\Lambda)=e^2\sum_{\alpha\beta\alpha'\beta',\gamma}
\delta_{\epsilon_{\alpha\beta},\epsilon_{\alpha'\beta'}}G_-^{\alpha\beta
\alpha'\beta'}(\vec X_{\beta\gamma}a^+_\alpha a_\gamma
a^+_{\beta'}a_{\alpha'} -\vec X_{\gamma\alpha}a^+_\gamma a_\beta
a^+_{\beta'}a_{\alpha'})\label{96}
\end{equation}
We consider here a situation of zero temperature and we compute
the mean value of $L_1(\vec X_\Lambda)$ on a Fock $N$--particle
state $\psi_I$:
\begin{equation}
\psi_I=a^+_{i_1}\dots a^+_{i_{N_I}}\psi_0, \quad \quad i_k\neq
i_l, \forall k\neq l\label{97}
\end{equation}
where $I$ is a set of possible quantum numbers, $N_I$ is the
number of electrons in $I$ and $\psi_0$ is the vacuum vector of
the fermionic operators, $a_\alpha \psi_0=0$ for all $\alpha$. The
order of the elements of $I$ is important to fix uniquely the
phase of  $\psi_I$. Equation (\ref{95}) gives now
\begin{equation}
\langle\psi_I,\vec
J_\Lambda(t)\psi_I\rangle\mid_{t=0}=\alpha_c\langle\psi_I,L(\vec
X_\Lambda) \psi_I\rangle\label{98}
\end{equation}
Introducing now the characteristic function of the set $I$,
\begin{equation}
\chi_I(\alpha)=\cases{ 1\hbox{ if }\alpha\in I\cr 0\hbox{ if
}\alpha\notin I,\cr}\label{99}
\end{equation}
we get
\begin{equation}
\langle a_\gamma^\dagger a_\alpha\psi_I,a_{\beta'}^\dagger
a_{\alpha'}\psi_I
\rangle=\delta_{\alpha\gamma}\delta_{\alpha'\beta'}\chi_I(\alpha)
\chi_I
(\alpha')+\delta_{\alpha\alpha'}\delta_{\gamma\beta'}\chi_I(\alpha)(1-\chi_I
(\gamma)), \label{100}
\end{equation}
see \cite{bagacc} for the details. Using this equality, together
with
\begin{equation}
\delta_{\varepsilon_{\alpha\beta},\varepsilon_{\alpha'\alpha'}}=\delta_{\varepsilon_{\alpha},
\varepsilon_{\beta}}\qquad\qquad\delta_{\varepsilon_{\alpha\beta},\varepsilon_{\alpha\beta'}}=
\delta_{\varepsilon_\beta,\varepsilon_{\beta'}}\label{101}
\end{equation}
we find that the average current is proportional to
\begin{equation}
\langle L(\vec X_\Lambda)\rangle_{\psi_I}={\cal L}_1(\vec
X_\Lambda)+{\cal L}_2(\vec X_\Lambda)\label{102}\end{equation}
where we isolate two contributions of different structure:
\begin{equation}
{\cal L}_1(\vec
X_\Lambda)=e^2\sum_{\alpha\beta\alpha'}\delta_{\varepsilon_\alpha,
\varepsilon_\beta} \{\chi_I(\alpha)-\chi_I(\beta)\}\chi_I(\alpha')
(\vec
X_{\alpha\beta}\overline{G_-^{\alpha\beta\alpha'\alpha'}}+\vec
X_{\beta\alpha}G_-^{\alpha\beta\alpha'\alpha'}), \label{103a}
\end{equation}
$${\cal L}_2(\vec X_\Lambda)=e^2\sum_{\alpha\beta\beta'}\delta_{\varepsilon_\beta,
\varepsilon_{\beta'}}\{\vec
X_{\beta\beta'}[G_-^{\alpha\beta\alpha\beta'}
\chi_I(\alpha)(1-\chi_I(\beta'))
-\overline{G_-^{\beta\alpha\beta'\alpha}}\chi_I(\beta')(1-\chi_I(\alpha))]-$$
\begin{equation}
-\vec X_{\beta'\beta}[G_-^{\beta\alpha\beta'\alpha}\chi_I(\beta')
(1-\chi_I(\alpha))
-\overline{G_-^{\alpha\beta\alpha\beta'}}\chi_I(\alpha)
(1-\chi_I(\beta'))]\}.\label{103b}
\end{equation}
\bigskip

Using equations (\ref{94a}), (\ref{94b}) for $X^{(i)}_{\gamma\mu}$
we are able to obtain ${\cal L}_1(X^{(i)}_\Lambda)$ and ${\cal
L}_2(X^{(i)}_\Lambda)$ for $i=1,2$. First of all we can show that,
even if ${\cal L}_1(X^{(1)}_\Lambda)$ is not zero, nevertheless it
does not depend on the electric field. Therefore
\begin{equation}
{\partial\over \partial E}{\cal L}_1(X^{(1)}_\Lambda)=0\label{104}
\end{equation}
Secondly, the computation of ${\cal L}_2(X^{(1)}_\Lambda)$ gives
rise to an interesting phenomenon: due to the definition of
$X^{(1)}_{\gamma\mu}$, the sum in (\ref{103b}) is different from
zero only if $p_\beta\neq p_{\beta'}$. Moreover, we also must have
$\varepsilon_\beta=\varepsilon_{\beta'}$, that is
\begin{equation}
n_\beta-n_{\beta'}={2\pi eE\over
m\omega^2L_x}\,(p_{\beta'}-p_\beta)\label{105}
\end{equation}
This equality can be satisfied in two different ways: let us
denote ${\cal R}$ the set of all possible quotients of the form
$(n_\beta-n_{\beta'})/(p_{\beta'}-p_\beta)$. This set, in
principle, coincides with the set of the rational numbers.
Therefore $0\in{\cal R}$. Then
\begin{itemize}
\item[1)] if ${2\pi eE\over m\omega^2L_x}$ is not in ${\cal R}$,
(\ref{105}) can be satisfied only if $\beta=\beta'$. But this
condition implies in particular that $p_\beta= p_{\beta'}$, and we
know already that whenever this condition holds, then
$X_{\beta\beta'}^{(1)}=0$, so that ${\cal L}_2
(X^{(1)}_\Lambda)=0$. \item[2)] If ${2\pi eE\over m\omega^2L_x}$
is in ${\cal R}$, then we have two possibilities: the first one is
again
$$\beta=\beta'$$
which, as we have just shown, does not contribute to ${\cal L}_2
(X^{(1)}_\Lambda)$. The second is
\begin{equation}
{n_\beta-n_{\beta'}\over p_{\beta'}-p_\beta}\,={2\pi eE\over
m\omega^2L_x}\label{107}
\end{equation}
which gives a non trivial contribution to the current.

Therefore, we can state the following
\end{itemize}\bigskip

\noindent{\sc Proposition}. {\sl In the context of our model
 there exists a set of rational numbers ${\cal R}$ with
the following property: if the electric and the magnetic fields
are such that  the quotient
$${2\pi eE\over m\omega^2L_x}$$
does not belong to ${\cal R}$ then
$$\langle J^{(1)}_\Lambda(t)\rangle_{\psi_I}=0.$$
}
\bigskip

On the other hand, if condition (\ref{107}) is satisfied, we can
conclude that the sum
$\sum_{\alpha\beta\beta'}\delta_{\varepsilon_\beta,\varepsilon_{\beta'}}
(\dots)$ in (\ref{103b}) can be replaced by
\begin{equation}
\sum_{\alpha\beta\beta'}\delta_{\varepsilon_\beta,\varepsilon_{\beta'}}
(\dots)=\sum_{\alpha}{\sum_{\beta\beta'}}'(\dots)\label{108}
\end{equation}
where $\sum_\alpha\sum'_{\beta\beta}$ means that the sum is
extended to all the $\alpha$ and to those $\beta$ and $\beta'$
with $p_\beta\neq p_{\beta'}$ satisfying (\ref{107}) (which
automatically implies that $\varepsilon_\beta=
\varepsilon_{\beta'}$).

Since, as it is easily seen,
$g_{\alpha\beta}(k)\overline{g_{\alpha'\beta'}(\underline k)}$
does not depend on $\vec E$, we find that
\begin{equation}
{\partial\over \partial
E}G^{\alpha\beta\alpha'\beta'}_-=-i{he\over m\omega
L_x}\,(p_\alpha-p_\beta)\Lambda^{\alpha\beta\alpha'\beta'}_-\label{109}
\end{equation}
where
\begin{equation}
\Lambda^{\alpha\beta\alpha'\beta'}_-=\int^0_{-\infty}d\tau \int d
kg_{\alpha\beta}(\underline
k)\overline{g_{\alpha'\beta'}(\underline
k)}e^{i\tau(\omega(\underline
k)-\varepsilon_{\alpha\beta})}\label{110}
\end{equation}
so that, using also (\ref{108}), we get
\begin{equation}
{\partial\over \partial E}{\cal L}_2(X_\Lambda^{(1)})={he\over
m\omega L_x}\Theta_x \label{(111)}
\end{equation}
where
$$\Theta_x:=\sum_{\alpha}{\sum_{\beta\beta'}}'(p_\beta-p_\alpha)
\tilde
x^{(1)}_{\beta\beta'}\{\chi_I(\alpha)(1-\chi_I(\beta'))\cdot
(\Lambda^{\alpha\beta\alpha\beta'}_-+\overline{\Lambda^{\alpha\beta
\alpha\beta'}_-})$$
\begin{equation}
-\chi_I(\beta')(1-\chi_I(\alpha))
(\Lambda^{\beta\alpha\beta'\alpha}_-+\overline{\Lambda^{\beta\alpha\beta'
\alpha}_-})\}\label{112}
\end{equation}
and
\begin{equation}
\tilde x^{(1)}_{\beta\beta'}=i \,
X^{(1)}_{\beta\beta'}\quad(\in{\bf R})\label{113}
\end{equation}
Therefore we conclude that
\begin{equation}
{\partial\over \partial E}\langle
J^{(1)}_\Lambda(t)\rangle_{\psi_I}={\alpha_c he^3\over m\omega
L_x}\, \Theta_x\label{114}
\end{equation}

Let us now compute the second component of the average current:
$\langle\psi_I, L(X^{(2)}_\Lambda)\psi_0\rangle={\cal
L}_1(X^{(2)}_\Lambda)+{\cal L}_2 (X^{(2)}_\Lambda)$.

The first contribution is easily shown, from (\ref{103a}) and
(\ref{94b}), to be identically zero, since
\begin{equation}
\delta_{\varepsilon_\alpha,\varepsilon_\beta}\delta_{p_\alpha
p_\beta} =\delta_{\alpha\beta}\label{115}
\end{equation}

On the contrary the second term, ${\cal L}_2(X^{(2)}_\Lambda)$, is
different from zero and it has an interesting expression: in fact,
due to the factor $\delta_{p_\mu, p_\gamma}$, the only non trivial
contributions in the sum
$\sum_{\beta\beta'}\delta_{\varepsilon_\beta,
\varepsilon_{\beta'}}$, in (\ref{103b}), are exactly those with
$\beta=\beta'$. Taking all this into account, we find that
\begin{equation}
{\cal
L}_2(X^{(2)}_\Lambda)=e^2\sum_{\alpha\beta}(y^{(p_\beta)}_0-y^{(p_\alpha)}_0)
\chi_I(\alpha)(1-\chi_I(\beta))
(G^{\alpha\beta\alpha\beta}_-+\overline{G^{\alpha\beta\alpha\beta}_-})\label{116}
\end{equation}
which is different from zero. Furthermore, using (\ref{109}), we
get
$${\partial\over \partial E}{\cal L}_2(X^{(2)}_\Lambda)=-2e^3\left({h\over m
\omega L_x}\right)^2 \Theta_y$$ were we have defined
\begin{equation}
\Theta_y
=\sum_{\alpha,\beta}(p_\alpha-p_\beta)^2\chi_I(\alpha)(1-\chi_I(\beta))\hbox{
Im }(\Lambda^{\alpha\beta\alpha\beta}_-)\label{117}
\end{equation}
and $\Lambda^{\alpha\beta\alpha\beta}_-$ is given by (\ref{110}).
If we call now
$$j_{x,E}={\partial \langle J^{(1)}_\Lambda(t)\rangle_{\psi_I}\over
\partial E} |_{t=0}=\alpha_c {\partial \langle L(X_\Lambda^{(1)})\rangle_{\psi_I}\over
\partial E}$$
$$j_{y,E}={\partial \langle J^{(2)}_\Lambda(t)\rangle_{\psi_I}\over
\partial E} |_{t=0}=\alpha_c {\partial \langle L(X_\Lambda^{(2)})\rangle_{\psi_I}\over
\partial E}\ ,$$
we obtain the conductivity tensor (see \cite{cha})
\begin{equation}
\sigma_{xx}=\sigma_{yy}=j_{y,E}, \quad\quad
\sigma_{xy}=-\sigma_{yx}= j_{x,E}\label{118}
\end{equation}
and the resistivity tensor
\begin{equation}
\rho_{xx}=\rho_{yy}={\sigma_{yy}\over\sigma_{yy}^2+\sigma_{xy}^2},
\quad\quad \rho_{xy}=-\rho_{yx}={\sigma_{xy}\over\sigma_{yy}^2+
\sigma_{xy}^2}\label{119}
\end{equation}
After minor computations we conclude that
\begin{equation}
\rho_{xy}=\cases{ 0\qquad\qquad\qquad\qquad\qquad\qquad\hbox{if
}{2\pi e E\over m\omega^2 L_x} \notin{\cal R}\cr {m\omega L_x\over
2 e^3h \alpha_c}{\Theta_x\over [\Theta_x^2+({h\over m\omega
L_x})^2\Theta_y^2]}\qquad\qquad\hbox{if }{2\pi e E\over m\omega^2
L_x}\in{\cal R}, \cr}\label{120}
\end{equation}
\begin{equation}
\rho_{xx}=\cases{ -({ m\omega L_x\over h})^2{1\over 2 \alpha_c
e^3\Theta_y}\qquad\qquad\qquad\qquad\hbox{if } {2\pi e E\over
m\omega^2 L_x}\notin{\cal R}\cr -{1\over 2
e^3\alpha_c}{\Theta_y\over [\Theta_x^2+({h\over m\omega L_x})^2
\Theta_y^2]}\qquad\qquad\qquad\hbox{if }{2\pi e E\over m\omega^2
L_x}\in{\cal R}\ ,\cr}\label{121}
\end{equation}

Let us now comment these results which are consequences of the
basic relation (\ref{107}). As it is evident from the formula
above, the fact that the {\it fine tuning condition\/} (FTC)
(${2\pi e E\over m\omega^2 L_x}\in{\cal R}$) is satisfied implies
that $\rho_{xy}\neq 0$, so that the resistivity tensor is
non-diagonal. Vice-versa, if the FTC is not satisfied, then
$\rho=\rho_{xx}1\!\!1$, $1\!\!1$ being the $2\times 2$ identity
matrix. This implies that, whenever the FTC holds, then the
$x$-component of the mean value of the density current operator is
in general different from zero, while it is necessarily zero if
the FTC is not satisfied.

If the physical system is prepared in such a way that ${2\pi e
E\over m\omega^2 L_x} \in{\cal R}$, then an experimental device
should be able to measure a non zero current along the $x$-axis.
Otherwise, this current should be zero whenever ${2\pi e E\over
m\omega^2 L_x}\notin{\cal R}$. A crucial point is now the
determination of the set ${\cal R}$, of rational numbers. From a
mathematical point of view, all the natural integers $n_\alpha$
and all the relative integer $p_\alpha$ are allowed. However
physics restricts the experimentally relevant values to a finite
set. In fact eigenstates corresponding to high values of
$n_\alpha$ and $p_\alpha$ are energetically not favoured because
the associated eigenenergies $\varepsilon_{n_\alpha p_\alpha}$ in
(\ref{15}) increases and the probabilities of finding an electron
in the corresponding eigenstate decrease (this is a generalization
of the standard argument which restrict the analysis of the
fractional QHE to the first few Landau levels). Moreover, high
positive values of $-p_\alpha$ are not compatible with the fact
that $H_0$ must be bounded from below, to be a {\em honest}
Hamiltonian.

Therefore, in formula (\ref{107}) not all the rational numbers are
physically allowed but only those compatible with the above
constraints. For this reason it is quite reasonable to expect that
the set ${\cal R}$ consists only of a {\it finite set\/} of
rational values. Of course, the determination of this set strongly
depends on the physics of the experimental setting.

Finally, let us remark that the sharp values of the magnetic field
involved in the FTC may be a consequence of the approximation
intrinsic in the stochastic limit procedure, which consists in
taking $\lambda\rightarrow 0$ and $t\rightarrow \infty$. In
intermediate regions ($\lambda> 0$ and $t<\infty$), it is not hard
to imagine that the $\delta$-function giving rise to the FTC
becomes a smoother function, and that real plateaux, closer to
what is observed in the QHE, appear.

Also, under special assumptions on the $B$-dependence of
$\Theta_x$ and $\Theta_y$, together with some reasonable physical
constraint on the value of the magnetic field, it is not difficult
to check that $\rho_{xx}$ has plateaux corresponding to the zeros
of $\rho_{xy}$ and that, outside of these plateaux, it grows
linearly with $B$.

\section{Laser Models}

In two recent papers, \cite{as,bs}, a dissipative laser model has
been introduced and analyzed in some details. In particular in
\cite{as} (AS in the following) the rigorous definition of the
unbounded generator of the model, which consists of a sum of a
free radiation and a free matter generator plus a matter-radiation
term, is given and the existence of the thermodynamical limit of
the dynamics of some macroscopic observables is deduced. Moreover,
the analysis of this dynamics shows that two phase transitions
occur in the model, depending on the value of a certain pumping
strength. In \cite{bs} the analysis has been continued paying
particular attention to the existence of the dynamics of the
microscopic observables, which are only those of the matter since,
in the thermodynamical limit, we proved that the field of the
radiation becomes classical. Also, the existence of a transient
has been proved and an {\em entropy principle} has been deduced.

On the other hand, in a series of papers \cite{dic,gra}
culminating with the fundamental work by Hepp and Lieb \cite{hl}
(HL in the following) many conservative models of matter
interacting with radiation were proposed. In particular, in
\cite{hl} the authors have introduced a model of an open system of
matter and of a single mode of radiation interacting among
themselves and with their own (bosonic) reservoirs, but, to
simplify the treatment, they have considered a simplified version
in which the matter bosonic reservoir is replaced by a fermionic
one. This is to avoid dealing with unbounded operators. This is
what they call the Dicke-Haken-Lax model (DHL model in the
following).

In \cite{as,bs} the relation between the AS model  and a many mode
version of the HL model is claimed: of course, since no reservoir
appear in the semigroup formulation  as given by \cite{as}, this
claim is reasonable but it is not clear the explicit way in which
HL should be related to AS. Here we will show that the relation
between the two models is provided by (a slightly modified version
of) the stochastic limit (SL). In particular, if we start with the
physical AS system (radiation and matter) and we introduce in a
natural way two reservoirs (one is not enough!) for the matter and
another reservoir for the radiation, then the SL of the
hamiltonian for the new system constructed in this way returns
back the original AS generator, under very reasonable hypotheses.
Moreover, the model which we have constructed {\em ad hoc} to get
this generator surprisingly coincides with the HL laser model,
\cite{hl}. This is the content of Subsection III.1, while in
Subsection III.2 we will consider the SL of the  DHL model,
\cite{hl,martin}. We will find that, even if the form of the
generator apparently differs from the one by AS, under certain
conditions on the coefficients which define the model, the
equations of motion for the observables of the matter-radiation
system coincide with the ones given in AS.

\vspace{2mm}

Let us discuss the main characteristics of the three physical
models which will be considered in this section. In particular, we
will only give the definition of the hamiltonians for the HL and
the DHL models and the expression of the generator for the AS
model, without even mentioning mathematical details like, for
instance, those related to the domain problem intrinsic with all
these models due to the presence of bosonic operators. We refer to
the original papers for these and further details which are not
relevant in this work.

We begin with the AS model.

This model is a dissipative quantum system, ${\Sigma}^{(N)},$
consisting of a chain of $2N+1$ identical two-level atoms
interacting with an $n-$mode radiation field, $n$ fixed and
finite. We build the model from its constituent parts starting
with the single atom. \vskip 0.2cm This is assumed to be a
two-state atom or spin, ${\Sigma}_{at}.$ Its algebra of
observables, ${\cal A}_{at},$ is that of the two-by-two matrices,
spanned by the Pauli matrices
$({\sigma}_{x},{\sigma}_{y},{\sigma}_{z})$ and the identity,
$\Id.$ They satisfy the relations \be
{\sigma}_{x}^{2}={\sigma}_{y}^{2}={\sigma}_{z}^{2}=\Id; \
{\sigma}_{x}{\sigma}_{y}=i{\sigma}_{z}, \ etc.\label{21II} \en We
define the spin raising and lowering operators
\be{\sigma}_{{\pm}}={1\over 2} ({\sigma}_{x}{\pm}i{\sigma}_{y}).
\label{2.2} \en We assume that the atom is coupled to a pump and a
sink, and that  its dynamics is given by a one-parameter semigroup
${\lbrace}T_{at}(t){\vert}t {\in}{\bf R}_{+}{\rbrace}$ of
completely positive, identity preserving contractions of ${\cal
A}_{at},$ whose generator, $L_{at},$ is of the following form. \be
L_{at}{\sigma}_{\pm}=-
({\gamma}_{1}{\mp}i{\epsilon}){\sigma}_{\pm}; \
L_{at}{\sigma}_{z}=-{\gamma}_{2}({\sigma}_{z}-
{\eta}I),\label{23II} \en where ${\epsilon}(>0)$ is the energy
difference between the ground and excited states of the atom, and
the ${\gamma}$'s and ${\eta}$ are constants whose values are
determined by the atomic coupling to the energy source and sink,
and are subject to the restrictions that \be
0<{\gamma}_{2}{\leq}2{\gamma}_{1}; \ -1{\leq}{\eta}{\leq}1.
\label{2.4} \en \vskip 0.2cm The matter consists of $2N+1$
non-interacting copies of ${\Sigma}_{at},$ located at the sites
$r=-N,. \ .,N$ of the one-dimensional lattice ${\bf Z}.$ Thus, at
each site, $r,$ there is a copy, ${\Sigma}_{r},$ of
${\Sigma}_{at},$ whose algebra of observables, ${\cal A}_{r},$ and
dynamical semigroup, $T_{r},$ are isomorphic with ${\cal A}_{at}$
and $T_{at},$ respectively. We denote by ${\sigma}_{r,u}$ the copy
of ${\sigma}_{u}$ at $r,$ for $u=x,y,z,{\pm}.$ \vskip 0.2cm We
define the algebra of observables, ${\cal A}^{(N)},$ and the
dynamical semigroup, $T_{mat}^{(N)},$ of the matter to be
${\otimes}_{r=-N}^{N}{\cal A}_{r}$ and
${\otimes}_{r=-N}^{N}T_{r},$ respectively. Thus, ${\cal A}^{(N)}$
is the algebra of linear transformations of ${\bf C}^{4N+2}.$ We
identify elements $A_{r}$ of ${\cal A}_{r}$ with those of ${\cal
A}^{(N)}$ given by their tensor products with the identity
operators attached to the remaining sites. Under this
identification, the commutant, ${\cal A}_{r}^{\prime},$ of ${\cal
A}_{r}$ is the tensor product ${\otimes}_{s{\neq}r}{\cal A}_{s}.$
The same identification will be implicitly assumed for the other
models. \vskip 0.2cm It follows from these specifications that the
generator, $L_{\rm mat}^{(N)},$ of $T_{mat}^{(N)}$ is given by the
formula \be L_{\rm mat}^{(N)}={\sum}_{l\in I_N}L_{l}, \label{2.5}
\en where  $I_N=\{-N,....,-1,0,1,..,N\}$. Here \be
L_{r}{\sigma}_{r,{\pm}}=-
({\gamma}_{1}{\mp}i{\epsilon}){\sigma}_{r,{\pm}}; \
L_{r}{\sigma}_{r,z}=-{\gamma}_{2}({\sigma}_{r,z}- {\eta}\Id);$$
$$\mbox{and }\ L_{r}(A_{r}A_{r}^{\prime})=(L_{r}A_{r})A_{r}^{\prime}
 \ {\forall}A_{r}{\in}{\cal A}_{r}, \
A_{r}^{\prime}{\in}{\cal A}_{r}^{\prime} \label{2.6} \en \vskip
0.3cm We assume, furthermore, that the radiation field consists of
$n(<{\infty})$ modes, represented by creation and destruction
operators ${\lbrace}a_{l}^{\star},a_{l}{\vert}l=0,. \
.,n-1{\rbrace}$ in a Fock-Hilbert space ${\cal H}_{rad}$ as
defined by the standard specifications that (a) these operators
satisfy the CCR, \be [a_{l},a_{m}^{\star}]={\delta}_{lm}\Id; \
[a_{l},a_{m}]=0, \label{27II} \en and (b) ${\cal H}_{rad}$
contains a (vacuum) vector ${\Phi}$, that is annihilated by each
of the $a$'s and is cyclic w.r.t. the algebra of polynomials in
the $a^{\star}$'s.

The formal generator of the semigroup $T_{rad}$ of the radiation
is \be L_{\rm rad}={\sum}_{l=0}^{n-1}\bigl(i{\omega}_{l}
[a_{l}^{\star}a_{l},.]+2{\kappa}_{l}a_{l}^{\star}(.)a_{l}-
{\kappa}_{l}{\lbrace}a_{l}^{\star}a_{l},.{\rbrace}\bigr),
\label{29} \en where ${\lbrace}.,.{\rbrace}$ denotes
anticommutator, and the frequencies, ${\omega}_{l},$ and the
damping constants, ${\kappa}_{l},$ are positive. We refer to
\cite{as} for a rigorous definition of $L_{\rm rad}$. \vskip 0.3cm
The composite (finite) system is simply the coupled system,
${\Sigma}^{(N)},$ comprising the matter and the radiation. We
assume that its algebra of observables, ${\cal B}^{(N)}$, is the
tensor product ${\cal A}^{(N)}{\otimes}{\cal R},$ where ${\cal R}$
is the $^{\star}-$algebra of polynomials in the $a$'s,
$a^{\star}$'s and the Weyl operators. Thus, ${\cal B}^{(N)},$ like
${\cal R},$ is an algebra of both bounded and unbounded operators
in the Hilbert space ${\cal H}^{(N)}:={\bf C}^{4N+2}{\otimes}{\cal
H}_{rad}$. We shall identify elements $A, \ R,$ of ${\cal
A}^{(N)}, \ {\cal R},$ with $A{\otimes}\Id_{rad}$ and
$\Id_{mat}{\otimes}R$,  respectively, with obvious notation.
\vskip 0.2cm We assume that the matter-radiation coupling is
dipolar and is given by the interaction Hamiltonian \be H_{\rm
int}^{(N)}={\sum}_{r\in I_N}
({\sigma}_{r,+}{\phi}_{r}^{(N)}+h.c.), \label{210III}\en where we
have introduced the so-called radiation field, ${\phi}^{(N)},$
whose value at the site $r$ is \be
{\phi}_{r}^{(N)}=-i(2N+1)^{-1/2}{\sum}_{l=0}^{n-1}{\lambda}_{l}
a_{l}{\exp}(2{\pi}ilr/n). \label{211III} \en Here the
${\lambda}$'s are real-valued, $N-$independent coupling constants.
\vskip 0.2cm Among the other results contained in \cite{as}, one
of the most relevant is that the map
$$L^{(N)}=L_{\rm mat}^{(N)}+L_{\rm rad}+i[H_{\rm int}^{(N)},.]$$
is really the generator of a $N$-depending semigroup, $T^{(N)}$,
regardless of the unbounded nature of both $L_{\rm rad}$ and
$H_{\rm int}^{(N)}$.

\vspace{4mm}

Let us now introduce the $n$-modes version of the HL model,
 \cite{hl}. The HL hamiltonian for
the $2N+1$ atoms and for the $n$ modes of the radiation can be
written as follows: \be H=H^{(S)}+H^{(R)}, \label{hl1} \en where
"S" refers to the system (radiation+matter) and "R" to the
reservoir. The hamiltonian of the system is \bea
& &H^{(S)}=\omega_R\sum_{j=0}^{n-1}a_j^\dagger a_j +\mu\sum_{l\in I_N}\sigma_{l,z}+\frac{\alpha}{\sqrt{2N+1}} \sum_{j=0}^{n-1}\sum_{l\in I_N}(\sigma_{l,+}a_je^{2{\pi}ijl/n}+\sigma_{l,-}a_j^\dagger e^{-2{\pi}ijl/n})+ \nonumber \\
& &+\frac{\beta}{\sqrt{2N+1}} \sum_{j=0}^{n-1}\sum_{l\in
I_N}(\sigma_{l,+}a_j^\dagger
e^{-2{\pi}ijl/n}+\sigma_{l,-}a_je^{2{\pi}ijl/n}), \label{hl2} \ena
\cite{baglaser}. Notice that the presence of $\beta$ means that we
are not restricting our model to the rotating wave approximation,
(RWA).

The hamiltonian for the reservoir contains two main contributions,
one related to the two reservoirs of the matter and one to the
reservoir of the radiation. We have: \be
H^{(R)}=H^{(P)}+\sum_{l\in I_N}H^{(A)}_l, \label{hl3} \en where
\be H^{(P)}=\sum_{j=0}^{n-1}\int dk \,
\omega_{r,j}(k)r_j(k)^\dagger r_j(k)+
\sqrt{\alpha}\sum_{j=0}^{n-1}(r_j^\dagger(\overline
g_j)a_j+r_j(g_j)a_j^\dagger), \label{hl4} \en and \be
H^{(A)}_l=\sum_{s=1}^2 \int dk \,\omega_{m_s}(k)m_{s,l}^\dagger(k)
m_{s,l}(k)+\sqrt{\alpha}(m_{1,l}^\dagger(\overline
h_1)\sigma_{l,-}+h.c.)+\sqrt{\alpha}(m_{2,l}^\dagger(\overline
h_2)\sigma_{l,+}+h.c.) \label{hl5} \en

We notice that:

1)  we are using the notation: $r_j(g_j)=\int dk \, r_j(k) g_j(k)$
and $r_j^\dagger(\overline g_j)=\int dk \,r_j^\dagger(k) \overline
g_j(k)$. Here $dk$ is a shortcut notation for $d\underline k^3$.

2) the functions $g_j$ and $h_{1,2}$ are introduced by HL to
regularize the bosonic fields $r_j(k)$ and $m_{(1,2),l}(k)$.

3) in this model two independent reservoirs, $m_{1,l}(k)$ and
$m_{2,l}(k)$, are introduced for (each atom of) the matter, while
only one, $r_j(k)$, is used for (each mode of) the radiation. This
result will be recovered also in our approach.

4) the coupling constant$\sqrt{\alpha}$ is written explicitly for
later convenience.

\vspace{3mm}

The role of each term of the hamiltonian above is evident. We
rewrite $H$ as a sum of a  free and of an interaction part, in the
following way: \be H=H_0+\sqrt{\alpha}H_I, \label{hl6} \en where

\be H_0=\omega_R\sum_{j=0}^{n-1}a_j^\dagger a_j +\mu\sum_{l\in
I_N}\sigma_{l,z}+\sum_{l\in I_N}\sum_{s=1}^2 \int dk
\omega_{m_s}(k)m_{s,l}^\dagger(k) m_{s,l}(k)+\sum_{j=0}^{n-1}\int
dk \omega_{r,j}(k)r_j(k)^\dagger r_j(k) \label{hl7} \en and \bea
H_I=\sum_{j=0}^{n-1}(r_j^\dagger(\overline g_j)a_j&&\hspace{-6mm}+r_j(g_j)a_j^\dagger)+\sum_{l\in I_N} [(m_{1,l}^\dagger(\overline h_1)\sigma_{l,-}+h.c.)+(m_{2,l}^\dagger(\overline h_2)\sigma_{l,+}+h.c.)]+\nonumber \\
&&+\frac{\sqrt{\alpha}}{\sqrt{2N+1}} \sum_{j=0}^{n-1}\sum_{l\in I_N}(\sigma_{l,+}a_je^{2{\pi}ijl/n}+\sigma_{l,-}a_j^\dagger e^{-2{\pi}ijl/n})+ \nonumber \\
&&+\frac{\beta}{\sqrt{\alpha(2N+1)}} \sum_{j=0}^{n-1}\sum_{l\in
I_N}(\sigma_{l,+}a_j^\dagger
e^{-2{\pi}ijl/n}+\sigma_{l,-}a_je^{2{\pi}ijl/n}). \label{hl8} \ena
The only non trivial commutation relations, which are different
from the ones already given in  (\ref{21II},\ref{27II}), are: \be
[r_j(k),r_l(k')^\dagger]=\delta_{j,l}\delta(k-k'), \hspace{5mm}
[m_{s,l}(k),m_{s',l'}^\dagger(k')]=\delta_{s,s'}\delta_{l,l'}\delta(k-k')
\label{hl9} \en

\vspace{4mm}

Finally, let us introduce the DHL model. The main difference,
\cite{martin}, consists in the use of a fermionic reservoir for
the matter, and for this reason the Pauli matrices of both AS and
HL are replaced by fermionic operators as described in details,
for instance, in \cite{martin}, \cite{baglaser}. Again we have \be
H=H_0+\lambda H_I, \label{dhl3} \en where, this time, \bea
&&H_0=\omega_R\sum_{j=0}^{n-1}a_j^\dagger a_j +\mu\sum_{l\in I_N}(b_{+,l}^\dagger b_{+,l}-b_{-,l}^\dagger b_{-,l})+\sum_{j=0}^{n-1}\int dk \,\omega_{r,j}(k)r_j(k)^\dagger r_j(k)+\nonumber\\
&&\hspace{-1cm}+\sum_{l\in I_N} \sum_{s=\pm}\int dk
\,\epsilon(k)(B_{s,l}^\dagger(k) B_{s,l}(k)+C_{s,l}^\dagger(k)
C_{s,l}(k)) \label{dhl4} \ena and \bea
&&H_I=\sum_{j=0}^{n-1}(r_j^\dagger(\overline
g_j)a+r_j(g_j)a_j^\dagger)+ \lambda {\sum}_{l\in I_N}
(\phi_{l}^{(N)}b_{+,l}^\dagger b_{-,l}+h.c.)+\nonumber \\
&&+\sum_{l\in I_N}\sum_{s=\pm}
[b_{s,l}^\dagger(B_{s,l}(g_{Bs})+C_{s,l}(g_{Cs}))+
(B_{s,l}^\dagger(g_{Bs})+C_{s,l}^\dagger(g_{Cs}))b_{s,l}].
\label{dhl5} \ena Here  $g_{B\pm}$ and $g_{C\pm}$ are real
function, and the $\{b_{\pm,l}^\sharp\}$ satisfy the following CAR
$\{b_{\pm,l}, b_{\pm,l}^\dagger\}=\1$ and they commutes when
localized at different lattice sites: $[b_{\pm,l},
b_{\pm,s}^\dagger]=0$ if $l\neq s$. .

The commutation rules for the radiation operators (system and
reservoir) coincide with the ones of the HL model. For what
concerns the matter operators (system and reservoirs) the first
remark is that any two operators localized at different lattice
sites commutes, as well as any operator of the radiation with any
observable of the matter. As for operators localized at the same
lattice site, the only non trivial anticommutators are \be
\{B_{\pm,l}(k),B_{\pm,l}^\dagger
(k')\}=\{C_{\pm,l}(k),C_{\pm,l}^\dagger (k')\}=\delta(k-k'),
\label{dhl6} \en  while all the others are zero. Finally, to
clarify the different roles between the $B$ and the $C$ fields it
is enough to consider their action on the ground state of the
reservoir $\varphi_0$: \be
r_j(k)\varphi_0=B_{\pm,l}(k)\varphi_0=C_{\pm,l}^\dagger(k)\varphi_0=0.
\label{dhl6bis} \en These equations, together with what has been
discussed, for instance, in \cite{martin}, show that $B$ is
responsible for the dissipation, while $C$ is the pump.

\subsection{Alli-Sewell versus  Hepp-Lieb}

We begin this subsection with a pedagogical note on the
single-mode single-atom version of the AS model. This will be
useful in order to show that two reservoirs must be introduced to
deal conveniently with the matter. After that we will consider the
full AS model and we will show that the hamiltonian which produces
the AS generator after considering its SL is nothing but the HL
hamiltonian in the RWA. We will finally comment that adding the
counter-rotating term (the one proportional to $\beta$ in
(\ref{hl2})) does not affect this result, since its contribution
disappear rigorously after the SL.

The starting point is given by the set of equations
(\ref{23II})-(\ref{211III}) restricted to $n=1$ and $N=0$, which
means only one mode of radiation and a single atom. With this
choice the phases  in $\phi_l^{(N)}$ disappear so that the
interaction hamiltonian (\ref{210III}) reduces to \be H_{\rm
int}=i(\sigma_-a^\dagger -h.c.), \label{31} \en and the total
generator is $L=L_{\rm mat}+L_{\rm rad}+i[H_{\rm int},.]$.

Let us suppose that the atom is coupled not only to the radiation
by means of $H_{\rm int}$, but also to a bosonic background $m(k)$
with the easiest possible dipolar interaction: \be
H_{Mm}=\sigma_+m(h)+h.c. \label{32} \en Of course this background
must have a free dynamics and the natural choice is \be
H_{0,m}=\int dk \,\omega_m (k) m^\dagger(k) m(k). \label{33} \en
For what concerns the radiation background the situation is
completely analogous: \be H_{0,r}=\int dk \,\omega_r (k)
r^\dagger(k) r(k), \hspace{2cm} H_{R,r}=ar^\dagger(\overline
g)+h.c. \label{34} \en are respectively the free hamiltonian and
the radiation-reservoir interaction. We take the complete
hamiltonian as simply the sum of all these contributions, with the
coupling constant $\lambda$ introduced as below: \bea
&&H=H_0+\lambda H_I=\{\mu \sigma_z+\omega_Ra^\dagger a+\int dk \, \omega_m (k) m^\dagger(k) m(k)+\int dk \,\omega_r (k) r^\dagger(k) r(k)\}+\nonumber \\
&&+\lambda \{(ar^\dagger(\overline
g)+h.c.)+(\sigma_+m(h)+h.c.)+\lambda i(\sigma_-a^\dagger -h.c.)\}.
\label{35} \ena Taking the SL of this model simply means, first of
all, considering the free evolution of the interaction
hamiltonian, $H_I(t)=e^{iH_0t}H_Ie^{-iH_0t}$. It is a simple
computation to obtain that, if $\omega_R=2\mu$, then \be
H_I(t)=(ar^\dagger(\overline
ge^{i(\omega_r-\omega_R)t})+h.c.)+(\sigma_+m(he^{i(2\mu-\omega_m)t})+h.c.)+\lambda
i(\sigma_-a^\dagger -h.c.). \label{37} \en In this case the SL
produces, as discussed in detail in \cite{baglaser}, the following
effective time-depending interaction hamiltonian: \be
H_I^{(sl)}(t)=(ar^\dagger_g(t)+h.c.)+(\sigma_+m_h(t)+h.c.)+
i(\sigma_-a^\dagger -h.c.), \label{38} \en where the dependence on
$\lambda$ disappears and the operators $r_g(t)$, $m_h(t)$ and
their hermitian conjugates satisfy the following commutation
relations for  $t> t'$, \be
[r_g(t),r_g^\dagger(t')]=\Gamma_-^{(g)}\delta (t-t'), \hspace{1cm}
[m_h(t),m_h^\dagger(t')]=\Gamma_-^{(h)}\delta (t-t'). \label{39}
\en Here we have defined the following complex quantities: \be
\Gamma_-^{(g)}=\int_{-\infty}^0d\tau \int dk
|g(k)|^2e^{-i(\omega_r(k)-\omega_R)\tau},
\hspace{4mm}\Gamma_-^{(h)}=\int_{-\infty}^0d\tau \int dk
|h(k)|^2e^{-i(2\mu-\omega_m(k))\tau}. \label{310} \en We want to
stress that the restriction $t>t'$ does not prevent to deduce the
commutation rules (\ref{313III}) below, which are the main
ingredient to compute the SL. However, the extension to $t<t'$ can
be easily obtained as discussed in \cite{book}. Of course the
functions $h$ and $g$ must be chosen in a such way that the
integrals above exist finite!

In order to obtain the generator of the model we introduce the
wave operator $U_t$ (in the interaction representation) which
satisfies the following operator differential equation: \be
\partial_tU_t=-iH_I^{(sl)}(t)U_t, \mbox{ with } \hspace{5mm} U_0=\Id.
\label{311III} \en We have already commented in Section II that,
at least for a large class of quantum mechanical models, the
equation above can be obtained as a suitable limit of differential
equations for a $\lambda$-depending wave operator, \cite{book}.
Analogously, $r_g(t)$ and $m_h(t)$ can be considered as the limit
(in the sense of the correlators) of the rescaled operators
$\frac{1}{\lambda}r( g e^{-i(\omega_r-\omega_R)t/{\lambda^2}})$
and $\frac{1}{\lambda}m(he^{i(2\mu-\omega_m)t/{\lambda^2}})$. It
is not surprising, therefore, that not only the operators but also
the vectors of the Hilbert space of the theory are affected by the
limiting procedure $\lambda\rightarrow 0$. In particular, the
vacuum $\eta_0$ for the operators $r_g$ and $m_h$,
$m_h(t)\eta_0=r_g(t)\eta_0=0$, does not coincide with the vacuum
$\varphi_0$ for $m(k)$ and $r(k)$,
$r(k)\varphi_0=m(k)\varphi_0=0$, see \cite{book} for more details.

Equation (\ref{311III}) above can be rewritten in the more
convenient form \be U_t=\Id-i\int_0^tH_I^{sl}(t')U_{t'}dt',
\label{312} \en which is used, together with the time consecutive
principle, \cite{book},  to obtain the following useful
commutation rules \be [r_g(t),U_t]=-i\Gamma_-^{(g)}aU_t,
\hspace{1cm}[m_h(t),U_t]=-i\Gamma_-^{(h)}\sigma_-U_t.
\label{313III} \en If we define the flow of a given observabe $X$
of the system as $j_t(X)=U_t^\dagger (X\otimes\Id_R)U_t$, the
generator is simply obtained by considering the expectation value
of $\partial_t j_t(X)$ on a vector state
$\eta_0^{(\xi)}=\eta_0\otimes \xi$, where  $\xi$ is a generic
state of the system. Using formulas (\ref{311III}), (\ref{313III})
and their hermitian conjugates, together with the properties of
the vacuum  $\eta_0$, the expression for the generator  follows by
identifying $L$ in the equation $\langle\partial_t
j_t(X)\rangle_{\eta_0^{(\xi)}}=\langle
j_t(L(X))\rangle_{\eta_0^{(\xi)}}$. The result is \bea
&&\hspace{-3cm}L(X)=L_1(X)+L_2(X)+L_3(X),\nonumber \\
&&\hspace{-3cm}L_1(X)=\Gamma_-^{(g)}[a^\dagger,X]a-\overline \Gamma_-^{(g)} a^\dagger [a,X],\hspace{.6cm}L_2(X)=\Gamma_-^{(h)}[\sigma_+,X]\sigma_--\overline \Gamma_-^{(h)} \sigma_+ [\sigma_-,X],\nonumber\\
&&\hspace{-3cm}L_3(X)=i^2[\sigma_-a^\dagger -\sigma_+a,X]
\label{314} \ena It is evident that both $L_1$ and $L_3$ can be
rewritten in the same form of the radiation and interaction terms
of the AS generator but this is not so, in general, for $L_2$
which has the form of the AS matter generator only if the pumping
parameter $\eta$ is equal to $-1$.

This is not very satisfactory and, how we will show in the
following, is a consequence of having introduced a single
reservoir for the atom. We will show that the existence of a
second reservoir  allows for the removal of the  constraint
$\eta=-1$ above.

With all of this in mind it is not difficult to produce an
hamiltonian which should produce the full AS generator for the
physical system with $2N+1$ atoms and  $n$ modes of radiation.
With respect to the one discussed above, it is enough to {\em
double} the number of reservoirs for the matter and to sum over
$l\in I_N$ for the matter and over $j=0,1,...,n-1$ for the modes.
The resulting hamiltonian is therefore \underline{necessarely}
very close to the HL one: \be H=H_0+\lambda H_I, \label{315} \en
with \be H_0=\omega_R\sum_{j=0}^{n-1}a_j^\dagger a_j
+\mu\sum_{l\in I_N}\sigma_{l,z}+\sum_{l\in I_N}\sum_{s=1}^2 \int
dk \,\omega_{m_s}(k)m_{s,l}^\dagger(k)
m_{s,l}(k)+\sum_{j=0}^{n-1}\int dk \,\omega_{r,j}(k)r_j(k)^\dagger
r_j(k) \label{316} \en and \bea
H_I=\sum_{j=0}^{n-1}(r_j^\dagger(\overline
g_j)a_j&&\hspace{-6mm}+r_j(g_j)a_j^\dagger)+\sum_{l\in I_N}
[(m_{1,l}^\dagger(\overline h_1)\sigma_{l,-}+h.c.)+
(m_{2,l}^\dagger(\overline h_2)\sigma_{l,+}+h.c.)]+\nonumber \\
&&+\lambda\sum_{l\in I_N}({\phi}_{l}^{(N)}\sigma_{l,+}+h.c.),
\label{317} \ena where the radiation field has been introduced in
(\ref{211III}). It is clear that, but for the RWA which we are
assuming here, there are not many other differences between this
hamiltonian and the one in (\ref{hl1})-(\ref{hl5}). It is worth
mentioning that $\lambda$ appears both as an overall coupling
constant, see (\ref{315}), and as a multiplying factor of
$\sum_{l\in I_N}({\phi}_{l}^{(N)}\sigma_{l,+}+h.c.)$ and plays the
same role as $\sqrt{\alpha}$ in the HL hamiltonian.   As for the
commutation rules they are quite natural: but for the spin
operators, which satisfy their own algebra, all the others
operators satisfy the CCR and commute whenever they refer to
different subsystems. In particular, for instance, all the
$m_{1,l}^\sharp(k)$ commute with all the $m_{2,l'}^\sharp(k')$,
for all $k,k'$ and $l,l'$.

The procedure to obtain the generator is the same as before: we
first compute $H_I(t)=e^{iH_0t}H_Ie^{-iH_0t}$, which enters in the
differential equation for the wave operator. Taking the limit
$\lambda\rightarrow 0$ of the mean value in the vector state
defined by $\varphi_0^{(\xi)}=\varphi_0 \otimes \xi$ of the first
non trivial approximation of the rescaled version of $U_t$ we
deduce the form of an effective hamiltonian, $H_I^{(sl)}(t)$,
which is simply \be
H_I^{(sl)}(t)=\sum_{j=0}^{n-1}(a_jr^\dagger_{g,j}(t)+h.c.)+\sum_{l\in
I_N}(\sigma_{l,+}m_{1,l}(t)+h.c.)+\sum_{l\in
I_N}(\sigma_{l,-}m_{2,l}(t)+h.c.)+\sum_{l\in
I_N}({\phi}_{l}^{(N)}\sigma_{l,+}+h.c.). \label{318} \en Again, we
are assuming that $\omega_R=2\mu$, which is crucial in order not
to have a time dependence in the last term of $H_I^{(sl)}(t)$ in
(\ref{318}).

The only non trivial commutation rules for $t>t'$ for the {\em
new} operators are: \bea
&&[r_{g,j}(t),r_{g,j'}^\dagger(t')]=\Gamma_{-,j}^{(g)}\delta_{j,j'}\delta (t-t'), \nonumber\\
&&[m_{1,l}(t),m_{1,l'}^\dagger(t')]=\Gamma_-^{(h_1)}\delta_{l,l'}\delta (t-t'),\\
\label{319}
&&[m_{2,l}(t),m_{2,l'}^\dagger(t')]=\Gamma_-^{(h_2)}\delta_{l,l'}\delta
(t-t'),\nonumber \ena where we have defined the following complex
quantities: \bea &&\Gamma_{-,j}^{(g)}=\int_{-\infty}^0d\tau \int
dk |g_j(k)|^2e^{i(\omega_{r,j}(k)-\omega_R)\tau},
\hspace{4mm}\Gamma_-^{(h_1)}=\int_{-\infty}^0d\tau \int dk |h_1(k)|^2e^{i(\omega_{m_1}(k)-\omega_R)\tau},\nonumber \\
&&\Gamma_-^{(h_2)}=\int_{-\infty}^0d\tau \int dk
|h_2(k)|^2e^{i(\omega_{m_2}(k)+\omega_R)\tau}. \label{320} \ena
From the commutation rules above and since $U_t=\Id-i\int
H_I^{(sl)}(t')U_{t'}dt'$, we get \be
[r_{g,j}(t),U_t]=-i\Gamma_{-,j}^{(g)}a_jU_t, \hspace{4mm}
[m_{1,l}(t),U_t]=-i\Gamma_{-}^{(h_1)}\sigma_{l,-}U_t, \hspace{4mm}
[m_{2,l}(t),U_t]=-i\Gamma_{-}^{(h_2)}\sigma_{l,+}U_t. \label{321}
\en The expression for the generator can be obtained as for the
$N=0$, $n=1$ model described before, that is computing the mean
value $\langle\partial_t j_t(X)\rangle_{\eta_0^{(\xi)}}$. Here, as
before, $j_t(X)$ is the flux of the system observable $X$,
$j_t(X)=U^\dagger_t (X\otimes\Id_R)U_t$, and $\eta_0$ is the
vacuum of the operators $r_{g,j}(t)$ and $m_{s,l}(t)$, $s=1,2$.
The computation gives the following result, which slightly
generalize the one in (\ref{314}): \bea
&&L(X)=L_1(X)+L_2(X)+L_3(X),\nonumber \\
&&L_1(X)=\sum_{j=0}^{n-1}(\Gamma_{-,j}^{(g)}[a_j^\dagger,X]a_j-\overline \Gamma_{-,j}^{(g)} a_j^\dagger [a_j,X]),\nonumber\\
&&L_2(X)=\sum_{l\in I_N}(\Gamma_-^{(h_1)}[\sigma_{+,l},X]\sigma_{-,l}-\overline \Gamma_-^{(h_1)} \sigma_{+,l} [\sigma_{-,l},X]+\Gamma_-^{(h_2)}[\sigma_{-,l},X]\sigma_{+,l}-\overline \Gamma_-^{(h_2)} \sigma_{-,l} [\sigma_{+,l},X],\nonumber\\
&&L_3(X)=i\sum_{l\in I_N}[({\phi}_{l}^{(N)}\sigma_{+,l}+h.c.),X].
\label{322} \ena It is not difficult to compare this generator
with the one proposed by AS, see formulas
((\ref{23II})-(\ref{211III})),
 and the conclusion is that the two
generators are exactly the same provided that the following
equalities are satisfied: \bea
&& \Im \Gamma_{-,j}^{(g)}=\omega_j, \hspace{3mm}\Re \Gamma_{-,j}^{(g)}=k_j, \hspace{3mm} \Re(\Gamma_{-}^{(h_1)}+\Gamma_{-}^{(h_2)})=\gamma_1, \hspace{3mm} \Im(\Gamma_{-}^{(h_1)}-\Gamma_{-}^{(h_2)})=\epsilon\nonumber\\
&& \Re\Gamma_{-}^{(h_1)}=\frac{1}{4}\gamma_2(1-\eta), \hspace{3mm}
\Re\Gamma_{-}^{(h_2)}=\frac{1}{4}\gamma_2(1+\eta). \label{323}
\ena Therefore, if we start with the HL hamiltonian, choosing the
regularizing functions in such a way that the equalities
(\ref{323}) are satisfied, the SL produces a generator of the
model which is exactly the one proposed in \cite{as,bs}, with the
extra minor constraint $\gamma_2=2\gamma_1$, which is a direct
consequence of (\ref{323}).

\vspace{3mm}

We conclude this subsection with a remark concerning the role of
the RWA and its relation with the SL. In particular, this is a
very good approximation after the SL is taken. To show why, we
first notice that adding a counter-rotating term (extending the
one in (\ref{hl8})) to the interaction hamiltonian $H_I$ in
(\ref{317}), considering the same coupling constant for both the
rotating and the counter-rotating term ($\beta=\alpha$), simply
means to add to $H_I$ in (\ref{317})  a contribution like
$\lambda\sum_{l\in I_N}({\phi}_{l}^{(N)}\sigma_{l,-}+h.c.)$. While
the rotating term, if $2\mu=\omega_R$, does not evolve freely, the
free time evolution of this other term is not trivial. However,
the differences with respect to the previous situation all
disappear rigorously after the SL, because these extra
contributions to the mean value of the wave operator go to zero
when $\lambda\rightarrow 0$, so that at the end the expression for
the generator is unchanged. This allows us to conclude that the
full HL hamiltonian is {\em equivalent} to the AS generator, where
the equivalence relation is provided by the SL.

\subsection{The SL of the DHL model}

We now consider the SL of the DHL model and we will get the
expression of the related generator showing that, under some
conditions on the quantities defining the model, the equations of
motion  do not differ from the ones in AS. The free evolved
interaction hamiltonian $H_I$ in (\ref{dhl5}) is, \bea
&&\hspace{-7mm}H_I(t)=e^{iH_0t}H_Ie^{-iH_0t}=\sum_{j=0}^{n-1}(a_jr_j^\dagger(\overline
g_je^{i(\omega_{r,j}-\omega_R)t})+h.c.)+ \lambda {\sum}_{l\in I_N}
(\phi_{l}^{(N)}b_{+,l}^\dagger b_{-,l}+h.c.)+\nonumber \\
&&\hspace{-7mm}+\sum_{l\in I_N} [b_{+,l}^\dagger(B_{+,l}(g_{B+}e^{it(\mu-\epsilon)})+C_{+,l}(g_{C+}e^{it(\mu-\epsilon)}))+ (B_{+,l}^\dagger(g_{B+}e^{-it(\mu-\epsilon)})+C_{+,l}^\dagger(g_{C+}e^{-it(\mu-\epsilon)}))b_{+,l}+\nonumber\\
&&\hspace{-7mm}+b_{-,l}^\dagger(B_{-,l}(g_{B-}e^{-it(\mu+\epsilon)})+C_{-,l}(g_{C-}e^{-it(\mu+\epsilon)}))+
(B_{-,l}^\dagger(g_{B-}e^{it(\mu+\epsilon)})+C_{-,l}^\dagger(g_{C-}e^{it(\mu+\epsilon)}))b_{-,l}].
\label{41} \ena Following the usual strategy we conclude that (the
rescaled version of)  the wave operator
$U_{\lambda}(t)=\Id-i\lambda\int_0^t H_I(t')U_{\lambda}(t')dt'$
converges for $\lambda\rightarrow 0$ to another operator, which we
still call a wave operator, satisfying the equation \be
U_t=\Id-i\int_0^t H_I^{(ls)}(t')U_{t'}dt', \mbox{ or, equivalently
}
\partial_t U_t=-i H_I^{(ls)}(t)U_{t}, \mbox{ with }\, U_0=\Id.
\label{42} \en Here $H_I^{(ls)}(t)$ is an effective time dependent
hamiltonian, found using the usual strategy, defined as \bea
H_I^{(ls)}(t)=&&\hspace{-7mm}\sum_{j=0}^{n-1}(a_jr_{g,j}^\dagger(t)+h.c.)+
{\sum}_{l\in I_N}
(\phi_{l}^{(N)}b_{+,l}^\dagger b_{-,l}+h.c.)+\sum_{l\in I_N} [b_{+,l}^\dagger(\beta_{+,l}(t)+\gamma_{+,l}(t))+ \nonumber \\
&&\hspace{-7mm}+(\beta_{+,l}^\dagger(t)+\gamma_{+,l}^\dagger(t))b_{+,l}+b_{-,l}^\dagger(\beta_{-,l}(t)+\gamma_{-,l}(t))+
(\beta_{-,l}^\dagger(t)+\gamma_{-,l}^\dagger(t))b_{-,l}].
\label{43} \ena The operators of the reservoir which appear in
$H_I^{(ls)}$ are the stochastic limit of the original (rescaled)
time evoluted operators of the reservoir and satisfy
(anti-)commutation relations which are related to those of the
original ones. In particular, after the SL, any two operators of
the matter (system and reservoirs) localized at different lattice
sites commutes, as well as any operator of the radiation with any
observable of the matter. As for operators localized at the same
lattice site, the only non trivial anticommutators are \be
\{\beta_{\pm,l}(t),\beta_{\pm,l}^\dagger
(t')\}=\delta(t-t')\Gamma_-^{(B\pm)}, \hspace{4mm}
\{\gamma_{\pm,l}(t),\gamma_{\pm,l}^\dagger
(t')\}=\delta(t-t')\Gamma_-^{(C\pm)}, \label{44} \en which should
be added to \be
[r_{g,j}(t),r_{g,j'}(t')]=\delta_{j,j'}\delta(t-t')\Gamma_{-,j}^{(g)}.
\label{45} \en In all these formulas the time ordering $t>t'$ has
to be understood and the following quantities are defined:

\bea &&\Gamma_{-,j}^{(g)}=\int_{-\infty}^0d\tau \int dk
\,|g_j(k)|^2e^{i(\omega_{r,j}(k)-\omega_R)\tau},
\hspace{4mm}\Gamma_-^{(B\pm)}=\int_{-\infty}^0d\tau \int dk \,(g_{B\pm}(k))^2e^{i(\epsilon(k)\mp \mu)\tau},\nonumber \\
&&\Gamma_-^{(C\pm)}=\int_{-\infty}^0d\tau \int dk
\,(g_{C\pm}(k))^2e^{-i(\epsilon(k)\mp \mu)\tau}. \label{45bis}
\ena
 We call now $\eta_0$ the vacuum of these limiting operators. We have
\be
r_{g,j}(t)\eta_0=\beta_{\pm,l}(t)\eta_0=\gamma_{\pm,l}^\dagger(t)\eta_0=0.
\label{46} \en Paying a little attention to the fact that here
commutators and anti-commutators simultaneously appear, we can
compute the commutators between the  operators $r_{g,j}(t)^\sharp,
\gamma_{\pm,l}^\sharp(t), \beta_{\pm,l}^\sharp(t)$  and the wave
operator $U_t$ by making use of (\ref{44},\ref{45}). We give here
only those commutation rules which are used in the computation of
the generator: \be [r_{g,j}(t),U_t]=-i\Gamma_{-,j}^{(g)}a_jU_t,
\hspace{4mm}
[\beta_{\pm,l}(t),U_t]=-i\Gamma_{-}^{(B\pm)}b_{\pm,l}U_t,
\hspace{4mm}
[\gamma_{\pm,l}^\dagger(t),U_t]=i\Gamma_{-}^{(C\pm)}b_{\pm,l}^\dagger
U_t. \label{47} \en Using the usual strategy, and restricting to
quadratic matter operators for technical reasons (this condition
can be avoided), \cite{baglaser}, we get  \bea
&&L(X)=L_1(X)+L_2(X)+L_3(X),\nonumber \\
&&L_1(X)=\sum_{j=0}^{n-1}(\Gamma_{-,j}^{(g)}[a_j^\dagger,X]a_j-\overline \Gamma_{-,j}^{(g)} a_j^\dagger [a_j,X]),\nonumber\\
&&L_2(X)=\sum_{l\in I_N}(\Gamma_-^{(B+)}[b_{+,l}^\dagger,X]b_{+,l}-\overline \Gamma_-^{(B+)} b_{+,l}^\dagger [b_{+,l},X]+\Gamma_-^{(C+)}[b_{+,l},X]b_{+,l}^\dagger-\overline \Gamma_-^{(C+)} b_{+,l} [b_{+,l}^\dagger,X]+\nonumber\\
&&\hspace{10mm}+\Gamma_-^{(B-)}[b_{-,l}^\dagger,X]b_{-,l}-\overline \Gamma_-^{(B-)} b_{-,l}^\dagger [b_{-,l},X]+\Gamma_-^{(C-)}[b_{-,l},X]b_{-,l}^\dagger-\overline \Gamma_-^{(C-)} b_{-,l} [b_{-,l}^\dagger,X]),\nonumber\\
&&L_3(X)=i\sum_{l\in I_N}[({\phi}_{l}^{(N)}b_{+,l}^\dagger
b_{-,l}+h.c.),X]. \label{48} \ena We see that the first and the
last terms exactly coincide with the analogous contributions of
the AS generator, but for a purely formal difference which is due
to the different matter variables which are used in the two
models. The second contribution, on the other hand, cannot be
easily compared with the free AS matter generator. What is
convenient, and sufficient, to get full insight about $L_2$, is to
compute its action on a basis of the local algebra, that is on
$b_{+,l}^\dagger b_{-,l}$ ($\equiv\sigma_{+,l}$) and on
$b_{+,l}^\dagger b_{+,l}-b_{-,l}^\dagger b_{-,l}$
($\equiv\sigma_{z,l}$), all the others being trivial or an easy
consequence of these ones. It is not hard to find the result: \bea
&&L_2(b_{+,l}^\dagger b_{-,l})=-b_{+,l}^\dagger b_{-,l}(\Re[\Gamma_-^{(B+)}+\Gamma_-^{(B-)}+\Gamma_-^{(C+)}+\Gamma_-^{(C-)}]-\nonumber\\
&&-i\Im[\Gamma_-^{(B+)}-\Gamma_-^{(B-)}-\Gamma_-^{(C+)}+\Gamma_-^{(C-)}]),\nonumber \\
&&L_2(b_{+,l}^\dagger b_{+,l}-b_{-,l}^\dagger b_{-,l})=2(-b_{+,l}^\dagger b_{+,l}(\Re[\Gamma_-^{(B+)}+\Gamma_-^{(C+)})+\Re \Gamma_-^{(C+)}+\nonumber\\
&&+b_{-,l}^\dagger
b_{-,l}(\Re[\Gamma_-^{(B-)}+\Gamma_-^{(C-)})-\Re \Gamma_-^{(C-)}).
\label{49} \ena The equation for $\sigma_{+,l}$ is recovered
without any problem, modulo some identifications
($\Re[\Gamma_-^{(B+)}+\Gamma_-^{(B-)}+\Gamma_-^{(C+)}+\Gamma_-^{(C-)}]=\gamma_1$,
...), while to recover the equation for $\sigma_{z,l}$ it is
necessary to choose properly the regularizing functions which
define the different $\Gamma_-$. In particular we need to have the
following equality fulfilled: \be
\Re(\Gamma_-^{(B+)}+\Gamma_-^{(C+)})=\Re(\Gamma_-^{(B-)}+\Gamma_-^{(C-)}).
\label{410} \en Under this condition we can conclude that the SL
of the DHL model produces the same differential equations as the
AS generator, as already happened for the HL model. It is also
easy to check that, as a consequence of our approach, we must have
$\gamma_1=\gamma_2$ in the generator we obtain. Of course this
result is not surprising since already in the HL paper, \cite{hl},
the fact that the two models are quite close (under some aspects)
was pointed out. Here we have learned also that the SL of both
these models, at least under some conditions, give rise to the
same dynamical behavior.

\section{The open BCS model}

In this section we review some results on the Open BCS model,
originally introduced in \cite{bm,martin}, obtained using the SLA,
\cite{bagbcs}.

The main outcome  is that the same values of the critical
temperature and of the order parameters can be found using the
SLA, in a significantly simpler way. This simplification allows us
to focus our attention on some aspects of the model which could
appear not so clearly using the original approach.

\subsection{The Physicals Model and its stochastic limit}

Our model consists of two main ingredients, the {\em system},
which is described by spin variables, and the {\em reservoir},
which is given in terms of bosonic operators. We refer to Section
III for the definition of the relevant operator algebras. The
system is contained in a box of volume $V=L^3$, with $N$ lattice
sites. We define, following \cite{bm,martin} \be
H_N^{(sys)}=\tilde\epsilon
\sum_{j=1}^N\sigma_j^0-\frac{g}{N}\sum_{i,j=1}^N\sigma_i^+\sigma_j^-,
\label{21} \en where the indexes $i,j$ represent the discrete
values of the momentum that an electron in a fixed volume can
have, $\sigma_j^+$ creates a Cooper pair with given momentum while
$\sigma_j^-$ annihilates the same pair, $\tilde\epsilon$ is the
energy of a single electron and $-g<0$ is the interaction close to
the Fermi surface. As we can see, only the $\pm$ component of the
spin, that is the $x,y$ components, have a mean field interaction,
while the $z$ component interacts with a constant external
magnetic field. The algebra of the Pauli matrices is given by \be
[\sigma_i^+,\sigma_j^-]=\delta_{ij}\sigma_i^0,\hspace{1cm}
[\sigma_i^\pm,\sigma_j^0]=\mp 2\delta_{ij}\sigma_i^\pm.
\label{22}\en We will use the following realization of these
matrices:
$$
\sigma^0\equiv\sigma^z=\left(
\begin{array}{cc}
1 & 0   \\
0 & -1  \\
\end{array}
\right), \hspace{5mm} \sigma^+=\left(
\begin{array}{cc}
0 & 1   \\
0 & 0  \\
\end{array}
\right), \hspace{5mm} \sigma^-=\left(
\begin{array}{cc}
0 & 0   \\
1 & 0  \\
\end{array}
\right).
$$
If we now define the following operators, \ \be
S_N^\alpha=\frac{1}{N}\sum_{i=1}^N\sigma_i^\alpha, \hspace{5mm}
R_N=S_N^+S_N^-=R_N^\dagger, \label{23}\en $H_N^{(sys)}$ can be
simply written as $H_N^{(sys)}=N(\tilde\epsilon S_N^0-gR_N)$ and
it is easy to check that the following commutation rules hold:
$$
[S_N^0,R_N]=[H_N^{(sys)},R_N]=[H_N^{(sys)},S_N^0]=0,
$$
for any given $N>0$. It is also worth noticing that the intensive
operators $S_N^\alpha$ are all bounded by 1 in the operator norm,
and that the commutators $[S_N^\alpha,\sigma_j^\beta]$ go to zero
in norm as $\frac{1}{N}$ when $N\rightarrow\infty$, for all $j,
\alpha$ and $\beta$.

\vspace{3mm}

Our construction of the reservoir follows the same steps given in
\cite{martin}, but for the commutation rules. We introduce here as
many bosonic modes $a_{\vec p,j}$ as lattice sites are present in
$V$. This means that $j=1,2,...,N$. $\vec p$ is the value of the
momentum of the j-th boson which, if we impose periodic boundary
condition on the wave functions, has necessarily the form $\vec
p=\frac{2\pi}{L}\vec n$, where $\vec n=(n_1,n_2,n_3)$ with
$n_j\in\Z$. These operators satisfy the following CCR, \be
[a_{\vec p,i},a_{\vec q,j}]=[a_{\vec p,i}^\dagger,a_{\vec
q,j}^\dagger]=0, \hspace{5mm} [a_{\vec p,i},a_{\vec
q,j}^\dagger]=\delta_{ij}\delta_{\vec p\,\vec q} \label{24}\en and
their free dynamics is given by \be
H_N^{(res)}=\sum_{j=1}^N\sum_{\vec p\in\Lambda_N}\epsilon_{\vec
p}\, a_{\vec p,j}^\dagger a_{\vec p,j}. \label{25} \en Here
$\Lambda_N$ is the set of values which $\vec p$ may take,
according to the previous remark: $\Lambda_N=\{\vec
p=\frac{2\pi}{L}\vec n,\, \vec n\in\Z^3\}$. It is useful to stress
that the energy of the different bosons is clearly independent of
the lattice site: $\epsilon_{\vec p}=\frac{\vec p^2}{2m} =
\frac{4\pi^2(n_1^2+n_2^2+n_3^2)}{2mL^2}$.

The  interaction is given by \be
H_N^{(I)}=\sum_{j=1}^N(\sigma_j^+a_j(f)+h.c.),\label{26}\en where
we have introduced $a_j(f)=\sum_{\vec p\in\Lambda_N}a_{\vec
p,j}f(\vec p)$, $f$ being a given test function which will be
asked to satisfy some extra conditions, see equation (\ref{221})
below and the related discussion.

The finite volume open system is now described by the following
hamiltonian, \be H_N=H_N^0+\lambda H_N^{(I)}, \mbox{ where }
H_N^0=H_N^{(sys)}+H_N^{(res)} \label{27}\en and $\lambda$ is the
coupling constant.

As usual, the first step in the SLA is the computation of the free
evolution of the interaction hamiltonian: \be
H_N^{(I)}(t)=e^{iH_N^0t}H_N^{(I)}e^{-iH_N^0t}=\sum_{j=1}^N(e^{iH_N^{(sys)}t}\sigma_j^+
e^{-iH_N^{(sys)}t}e^{iH_N^{(res)}t}a_j(f)
e^{-iH_N^{(res)}t}+h.c.). \label{28}\en The computation of the
part of the reservoir is trivial and produces
$$e^{iH_N^{(res)}t}a_j(f) e^{-iH_N^{(res)}t}=a_j(fe^{-it\epsilon
}),$$ where $a_j(fe^{-it\epsilon })=\sum_{\vec
p\in\Lambda_N}a_{\vec p,j}f(\vec p)e^{-it\epsilon_{\vec p}}$. This
is an easy consequence of the CCR (\ref{24}). The free evolution
of the spin operators is more difficult and its expression can be
found in \cite{bm,martin}, for instance, where it is shown how to
obtain the time evolution in a {\em semiclassical} approximation,
i.e., when the free time evolution of the intensive operators
$S_N^\alpha$ are replaced by their limits  in a suitable topology,
\cite{bagmor} and \cite{bagbcs}.

The differential equations of motion for the spin variables are
\be \left\{
\begin{array}{ll}
\frac{d\sigma_j^+(t)}{dt} = 2i\tilde\epsilon \sigma_j^+(t)+igS_N^+(t)\sigma_j^0(t)  \\
\frac{d\sigma_j^0(t)}{dt} = 2ig(\sigma_j^+(t)S_N^-(t)-\sigma_j^-(t)S_N^+(t)). \\
\end{array}
\right. \label{29} \en where $\sigma_j^\alpha(t)=
e^{iH_N^{(sys)}t}\sigma_j^\alpha e^{-iH_N^{(sys)}t}$ and
$S_N^\alpha(t)=e^{iH_N^{(sys)}t}S_N^\alpha e^{-iH_N^{(sys)}t}=
\frac{1}{N}\sum_{j=1}^N e^{iH_N^{(sys)}t}\sigma_j^\alpha
e^{-iH_N^{(sys)}t}= \frac{1}{N}\sum_{j=1}^N \sigma_j^\alpha(t)$.

Let us now call $S^\alpha={\mathcal F}-strong
\lim_{N\rightarrow\infty}S^\alpha_N$, where $\mathcal F$ is a
suitable family of vectors. The proof of the existence of this
limit (together with all its powers) may be found in \cite{bagmor}
and references therein. We can now take the sum over $j=1,2,...,N$
of (both sides of) the equations in (\ref{29}), divide the result
by $N$, and then consider the ${\mathcal F}-strong
\lim_{N\rightarrow\infty}$ of the equations obtained in this way.
We find that $\dot S^0(t)=0$ and $\dot
S^+(t)=i(2\tilde\epsilon+gS^0(t))S^+(t)$. These equations can be
easily solved: $S^0(t)=S^0=(S^0)^\dagger$ and
$S^+(t)=S^+e^{i(2\tilde\epsilon+gS^0)t}$. Of course $S^-(t)=
(S^+(t))^\dagger$. The system (\ref{29}) becomes, if we replace
$S_N^\alpha(t)$ with its ${\mathcal F}-strong$ limit
$S^\alpha(t)$, \be \left\{
\begin{array}{ll}
\frac{d\sigma_j^+(t)}{dt} = 2i\tilde\epsilon \sigma_j^+(t)+igS^+(t)\sigma_j^0(t)  \\
\frac{d\sigma_j^0(t)}{dt} = 2ig(\sigma_j^+(t)S^-(t)-\sigma_j^-(t)S^+(t)). \\
\end{array}
\right. \label{29bis} \en This system is called the {\em
semiclassical  approximation} of (\ref{29}), and it can be
explicitly solved: \be \sigma_j^+(t)=e^{i\nu
t}\rho_0^j+e^{i(\nu+\omega) t}\rho_+^j+e^{i(\nu-\omega)
t}\rho_-^j, \label{210}\en where we have defined the following
operators \be \left\{
\begin{array}{ll}
\rho_0^j = \frac{g^2 S^+}{\omega^2}\left(2S^-\sigma_j^++S^0\sigma_j^0+2S^+\sigma_j^-\right)  \\
\rho_+^j = \frac{g
S^+}{\omega^2}\left(gS^-\frac{\omega-gS^0}{\omega+gS^0}\sigma_j^++
\frac{\omega-gS^0}{2}\sigma_j^0-gS^+\sigma_j^-\right) \\
\rho_-^j = \frac{g S^+}{\omega^2}
\left(gS^-\frac{\omega+gS^0}{\omega-gS^0}\sigma_j^+-
\frac{\omega+gS^0}{2}\sigma_j^0-gS^+\sigma_j^-\right), \\
\end{array}
\right. \label{211} \en and the following quantities \be
\omega=g\sqrt{(S^0)^2+4S^+S^-},\: \nu=2\tilde\epsilon+gS^0.
\label{212}\en Defining further \be \nu_\alpha(\vec
p)=\nu-\epsilon_{\vec p}+\alpha\omega, \label{212bis}\en
 where $\alpha$ takes the values $0$, $+$ and
$-$, the operator $H_N^{(I)}(t)$ in (\ref{28}) becomes \be
H_N^{(I)}(t)=\sum_{j=1}^N\sum_{\alpha=0,\pm}\left(\rho_\alpha^ja_j(fe^{it\nu_\alpha})+h.c\right).
\label{213}\en The next step in the SLA consists in computing the
following quantity \be
I_\lambda(t)=\left(-\frac{i}{\lambda}\right)^2\int_0^t dt_1
\int_0^{t_1}dt_2\,\omega_{tot}\left(H_N^{(I)}(\frac{t_1}{\lambda^2})H_N^{(I)}(\frac{t_2}{\lambda^2})\right),\label{214}\en
and its limit for $\lambda$ going to zero. Here the state
$\omega_{tot}$ is the following product state
$\omega_{tot}=\omega_{sys}\,\omega_\beta$, where $\omega_{sys}$ is
a generic state of the system, while $\omega_\beta$ is a state of
the reservoir, which we will take to be a KMS state corresponding
to an inverse temperature $\beta=\frac{1}{kT}$. It is convenient
here to use the so-called {\em canonical representation of thermal
states}, \cite{book}, which is sketched in the Appendix. Then we
introduce two sets of mutually commuting bosonic operators
$\{c_{\vec p,j}^{(\gamma)}\}$, $\gamma=a,b$, as follows: \be
a_{\vec p,j}=\sqrt{m(\vec p)}\,c_{\vec p,j}^{(a)}+\sqrt{n(\vec
p)}\,c_{\vec p,j}^{(b),\dagger},\label{215}\en where \be m(\vec
p)=\omega_\beta(a_{\vec p,j}a_{\vec
p,j}^\dagger)=\frac{1}{1-e^{-\beta\epsilon_{\vec p}}},
\hspace{1cm}n(\vec p)=\omega_\beta(a_{\vec p,j}^\dagger a_{\vec
p,j})=\frac{e^{-\beta\epsilon_{\vec p}}}{1-e^{-\beta\epsilon_{\vec
p}}}.\label{215bis}\en The operators $c_{\vec p,j}^{(\alpha)}$
satisfy the following commutation rules \be [c_{\vec
p,j}^{(\alpha)},{c_{\vec
q,k}^{(\gamma)}}^\dagger]=\delta_{jk}\delta_{\vec p\,\vec
q}\delta_{\alpha\gamma}, \label{216}\en while all the other
commutators are trivial. Furthermore, we introduce  the vacuum of
the operators $c_{\vec p,j}^{(\alpha)}$, $\Phi_0$: \be c_{\vec
p,j}^{(\alpha)}\Phi_0=0, \hspace{1cm}\forall \vec p\in\Lambda_N,\:
j=1,..N,\: \alpha=a,b. \label{219}\en Finally, if we define
$f_m(\vec p)=\sqrt{m(\vec p)}f(\vec p)$ and $f_n(\vec
p)=\sqrt{n(\vec p)}f(\vec p)$, we get \be
a_j(fe^{it\nu_\alpha})=c_j^{(a)}(f_me^{it\nu_\alpha}) +
{c_j^{(b)}}^\dagger(f_ne^{it\nu_\alpha}),\label{217}\en using the
usual  notation for $c_j^{(a)}(g)$ and ${c_j^{(b)}}^\dagger(g)$.
Therefore we have \be
H_N^{(I)}(t)=\sum_{j=1}^N\sum_{\alpha=0,\pm}\left\{\rho_\alpha^j\left(c_j^{(a)}(f_me^{it\nu_\alpha})
+ {c_j^{(b)}}^\dagger(f_ne^{it\nu_\alpha})\right)+h.c\right\},
\label{218}\en and the KMS state $\omega_\beta$ can be represented
as the following vector state, as in a GNS-like representation:
\be \omega_\beta(X_r)=\langle\Phi_0, X_r \Phi_0\rangle,
\label{220}\en for any observable of the reservoir, $X_r$, since
$\omega_\beta$ is a gaussian state, \cite{book}. This fact,
together with (\ref{219}) and with the commutation rules
(\ref{216}), simplifies the computation of the two point function
$\omega_{tot}\left(H_N^{(I)}(\frac{t_1}{\lambda^2})H_N^{(I)}(\frac{t_2}{\lambda^2})\right)$,
which produces
$$
\omega_{tot}\left(H_N^{(I)}(\frac{t_1}{\lambda^2})H_N^{(I)}(\frac{t_2}{\lambda^2})\right)=\sum_{j=1}^N
\sum_{\alpha,\beta=0,\pm}\sum_{\vec
p\in\Lambda_N}\{\omega_{sys}(\rho_\alpha^j{\rho_\beta^j}^\dagger)|f_m(\vec
p)|^2 e^{i\frac{t_1}{\lambda^2}\nu_\alpha(\vec p)}
e^{-i\frac{t_2}{\lambda^2}\nu_\beta(\vec p)}+$$ \vspace{-4mm}
$$+\, \omega_{sys}({\rho_\alpha^j}^\dagger \rho_\beta^j)|f_n(\vec
p)|^2 e^{-i\frac{t_1}{\lambda^2}\nu_\alpha(\vec p)}
e^{+i\frac{t_2}{\lambda^2}\nu_\beta(\vec p)} \}.
$$
Since we are interested to the limit $\lambda\rightarrow 0$ of
$I_\lambda(t)$ we need to impose some conditions on the test
function $f(\vec p)$, \cite{book}. In particular, we will require
that the following integral exists finite: \be
\int_{-\infty}^0\,d\tau \sum_{\vec p\in\Lambda_N}|f_r(\vec
p)|^2e^{\pm i\tau\nu_\alpha(\vec p)}<\infty,\label{221}\en where
$f_r(\vec p)$ is $f_m(\vec p)$ or $f_n(\vec p)$ and
$\nu_\alpha(\vec p)$ is given in (\ref{212bis}). Under this
assumption we find that \be I(t)=\lim_{\lambda\rightarrow
0}I_\lambda(t)=-t\sum_{j=1}^N\sum_{\alpha=0,\pm}\left\{\omega_{sys}
(\rho_\alpha^j{\rho_\alpha^j}^\dagger)\Gamma_\alpha^{(a)}+
\omega_{sys}({\rho_\alpha^j}^\dagger
\rho_\alpha^j)\Gamma_\alpha^{(b)}\right\}, \label{222}\en where
the two complex quantities \be
\Gamma_\alpha^{(a)}=\int_{-\infty}^0\,d\tau \sum_{\vec
p\in\Lambda_N}|f_m(\vec p)|^2e^{- i\tau\nu_\alpha(\vec p)},
\hspace{5mm} \Gamma_\alpha^{(b)}=\int_{-\infty}^0\,d\tau
\sum_{\vec p\in\Lambda_N}|f_n(\vec p)|^2e^{i\tau\nu_\alpha(\vec
p)} \label{223}\en both exist because of the assumption
(\ref{221}).

To this  same result we could also arrive starting with the
following {\em stochastic limit hamiltonian} \be
H_N^{(sl)}(t)=\sum_{j=1}^N\sum_{\alpha=0,\pm}\left\{\rho_\alpha^j\left(c_{\alpha
j}^{(a)}(t) + {c_{\alpha j}^{(b)}}^\dagger(t)\right)+h.c\right\},
\label{224}\en where the operators $c_{\alpha j}^{(\gamma)}(t)$
 satisfy the following commutation rule, \be
[c_{\alpha j}^{(\gamma)}(t),{c_{\beta
k}^{(\mu)}}^\dagger(t')]=\delta_{jk}\,\delta_{\alpha
\beta}\,\delta_{\gamma \mu} \delta(t-t')\Gamma_\alpha^{(\gamma)},
\hspace{1.4cm}\mbox{for } t>t'. \label{225}\en

We mean that, as it is easily checked, the following quantity
$$J(t)=(-i)^2\int_0^t dt_1
\int_0^{t_1}dt_2\Omega_{tot}(H_N^{(sl)}(t_1)H_N^{(sl)}(t_2))$$
coincides with $I(t)$. Here $\Omega_{tot}= \omega_{sys}\,
\Omega=\omega_{sys}\, \langle\Psi_0,\,\Psi_0\rangle$, where
$\Psi_0$ is the vacuum of the operators $c_{\alpha
j}^{(\gamma)}(t)$: $c_{\alpha j}^{(\gamma)}(t)\Psi_0=0$ for all
$\alpha, j, \gamma$ and $t$, \cite{book}.

\vspace{2mm}

We now use $H_N^{(sl)}(t)$ to compute the generator of the theory.
 Let $X$ be an observable of the system and $\Id_r$
the identity of the reservoir. Its time evolution (after the
stochastic limit is taken) is $j_t(X\otimes \Id_r)=U_t^\dagger
(X\otimes \Id_r) U_t$, where $U_t$ is the wave operator satisfying
the following differential equation
$\partial_tU_t=-iH_N^{(sl)}(t)U_t$, whose adjoint is
$\partial_tU_t^\dagger=iU_t^\dagger H_N^{(sl)}(t)$.

Then we find
$$
\partial_tj_t(X\otimes\Id_r)=iU_t^\dagger[H_N^{(sl)}(t),X\otimes\Id_r]U_t=$$
\vspace{-6mm} $$=iU_t^\dagger\sum_{j=1}^N\sum_{\alpha=0,\pm}
\left\{[\rho_\alpha^j,X](c_{\alpha j}^{(a)}(t)+{c_{\alpha
j}^{(b)}}^\dagger(t))+ [{\rho_\alpha^j}^\dagger,X]({c_{\alpha
j}^{(a)}}^\dagger(t)+c_{\alpha j}^{(b)}(t)) \right\}U_t
$$
Using now the commutation rules \be [c_{\alpha
j}^{(a)}(t),U_t]=-i\int_0^t [c_{\alpha
j}^{(a)}(t),H_N^{(sl)}(t')]U_{t'}\,dt'= -i\int_0^t
({\rho_\alpha^j}^\dagger \Gamma_\alpha^{(a)}\delta
(t-t'))U_{t'}\,dt'=-i {\rho_\alpha^j}^\dagger \Gamma_\alpha^{(a)}
U_t\label{226}\en and \be [c_{\alpha j}^{(b)}(t),U_t]=-i
{\rho_\alpha^j} \Gamma_\alpha^{(b)} U_t, \label{227}\en and their
adjoints, we find that
$$
\partial_tj_t(X\otimes\Id_r)=i{\sum_{j=1}^N}\sum_{\alpha=0\pm}\mbox{\LARGE\{}\left(iU_t^\dagger
{\rho_\alpha^j}^\dagger \overline{\Gamma_\alpha^{(b)}}+{c_{\alpha
j}^{(b)}}^\dagger(t)U_t^\dagger\right)[\rho_\alpha^j,X]U_t + $$
\vspace{-3mm} $$\left(iU_t^\dagger {\rho_\alpha^j}
\overline{\Gamma_\alpha^{(a)}}+{c_{\alpha
j}^{(a)}}^\dagger(t)U_t^\dagger\right)[{\rho_\alpha^j}^\dagger,X]U_t+
U_t^\dagger [\rho_\alpha^j,X]\left(-i{\rho_\alpha^j}^\dagger
\Gamma_\alpha^{(a)} U_t+U_t c_{\alpha j}^{(a)}(t) \right)+
$$
\vspace{-3mm} $$+ U_t^\dagger
[{\rho_\alpha^j}^\dagger,X]\left(-i{\rho_\alpha^j}
\Gamma_\alpha^{(b)} U_t+U_t c_{\alpha j}^{(b)}(t)
\right)\mbox{\LARGE\}}
$$
which has to be computed on the state $\Omega_{tot}$. Therefore,
since the generator $L$ satisfies the equality
$\Omega_{tot}(\partial_tj_t(X\otimes\Id_r))=\Omega_{tot}(j_t(L(X)))$,
we  get \be
L(X)={\sum_{j=1}^N}\sum_{\alpha=0\pm}\left\{[\rho_\alpha^j,X]
{\rho_\alpha^j}^\dagger
\Gamma_\alpha^{(a)}+[{\rho_\alpha^j}^\dagger,X] {\rho_\alpha^j}
\Gamma_\alpha^{(b)}- \rho_\alpha^j [{\rho_\alpha^j}^\dagger,X]
\overline{\Gamma_\alpha^{(a)}}- {\rho_\alpha^j}^\dagger
[{\rho_\alpha^j},X]
\overline{\Gamma_\alpha^{(b)}}\right\}\label{228} \en This
expression can be made simpler if the observable $X$ is
self-adjoint ($X=X^\dagger$). In this case we have \be
L(X)=L_1(X)+L_2(X),\label{229}\en where \be
L_1(X)=\sum_{j=1}^N\sum_{\alpha=0\pm}\left\{[\rho_\alpha^j,X]
{\rho_\alpha^j}^\dagger \Gamma_\alpha^{(a)}+ h.c. \right\},
\hspace{4mm}
L_2(X)=\sum_{j=1}^N\sum_{\alpha=0\pm}\left\{[{\rho_\alpha^j}^\dagger,X]
{\rho_\alpha^j} \Gamma_\alpha^{(b)}+ h.c. \right\}.\label{230}\en

\subsection{The phase transition}

As discussed in \cite{bm,martin}, $S_N^0$ and $R_N$ are the
relevant variables whose dynamics must be considered to analyze
the phase structure of the model. These intensive operators are
both self-adjoint, so that we can use equations (\ref{229}) and
(\ref{230}) instead of (\ref{228}). As a matter of fact, in both
\cite{bm} and \cite{martin} these equations of motion are
considered only as an intermediate step to compute the equation
for $\Delta_N=\frac{1}{2}R_N^{1/2}$, which is called {\em the gap
operator}. We have shown in \cite{bagbcs} that the same
conclusions as in \cite{bm,martin} can be obtained without
introducing $\Delta_N$ but working directly with $R_N$ and
$S_N^0$.

First we focus on $L(S_N^0)=L_1(S_N^0)+L_2(S_N^0)$. We have, using
(\ref{22}), (\ref{23}) and (\ref{230})
$$L_1(S_N^0)=\frac{1}{N}\sum_{j=1}^N
L_1(\sigma_j^0)=\frac{1}{N}\sum_{j=1}^N\sum_{\alpha=0,\pm}\left\{[\rho_\alpha^j,
\sigma_j^0] {\rho_\alpha^j}^\dagger
\Gamma_\alpha^{(a)}+h.c.\right\}$$ whose limit in the ${\mathcal
F}-strong$ topology exists, \cite{bagmor}, and is given by
 \be L_1(S^0):={\mathcal F}-strong
\lim_{N\rightarrow\infty}
L_1(S^0_N)=-\frac{8g^4S^0(S^+S^-)^2}{\omega^3}\left\{\Re
\Gamma_+^{(a)}\frac{\omega-g}{(\omega+gS^0)^2}+ \Re
\Gamma_-^{(a)}\frac{\omega+g}{(\omega-gS^0)^2}\right\},
\label{32}\en where $\Re \Gamma_\pm^{(a)}$ indicates the real part
of $\Gamma_\pm^{(a)}$, \cite{bagbcs}.

The computation of $L_2(S^0):={\mathcal F}-strong
\lim_{N\rightarrow\infty} L_2(S^0_N)$ follows essentially the same
steps and produces \be
L_2(S^0)=-\frac{8g^4S^0(S^+S^-)^2}{\omega^3}\left\{\Re
\Gamma_+^{(b)}\frac{\omega+g}{(\omega+gS^0)^2}+ \Re
\Gamma_-^{(b)}\frac{\omega-g}{(\omega-gS^0)^2}\right\},
\label{33}\en \cite{bagbcs}, so that the final result is \be
L(S^0)=-\frac{8g^4S^0(S^+S^-)^2}{\omega^3}h(S^0,S^+S^-).\label{34}\en
Here we have introduced, for brevity, the function \be
h(S^0,S^+S^-)=\Re \Gamma_+^{(a)}\frac{\omega-g}{(\omega+gS^0)^2}+
\Re \Gamma_-^{(a)}\frac{\omega+g}{(\omega-gS^0)^2}+\Re
\Gamma_+^{(b)}\frac{\omega+g}{(\omega+gS^0)^2}+ \Re
\Gamma_-^{(b)}\frac{\omega-g}{(\omega-gS^0)^2}, \label{35}\en and
we have written explicitly the dependence of $h$  on
$S^+S^-={\mathcal F}-strong \lim_{N\rightarrow\infty}S_N^+S_N^-$,
see (\ref{212}).  It is interesting to observe that the same
function $h(S^0,S^+S^-)$ appears in the computation of
$L(S^+S^-):={\mathcal F}-strong \lim_{N\rightarrow\infty}
L(S_N^+S_N^-)$. Again, since $(S_N^+S_N^-)^\dagger=S_N^+S_N^-$, we
can use formulas (\ref{229}) and (\ref{230}). Here the
computations are significantly harder, but no difficulty of
principle arises. As a technical tool it is convenient to use  the
fact that, in the limit $N\rightarrow\infty$, all the intensive
operators commute with all the local operators of the system,
$\lim_{N\rightarrow\infty}[S_N^\alpha,\sigma_j^\beta]=0$, for all
$\alpha, \beta$ and $j$. Therefore we get \be
L(S^+S^-)=-\frac{16g^4(S^+S^-)^3}{\omega^3}h(S^0,S^+S^-).\label{36}\en

\vspace{3mm}

The  phase  structure of the model is now given by the right-hand
sides of equations (\ref{34}) and (\ref{36}), see
\cite{bm,martin}, and, in particular, from the zeros of the
functions \be f_1(x,y)=-\frac{8g^4xy^2}{\omega^3}h(x,y),
\hspace{5mm}f_2(x,y)=-\frac{16g^4y^3}{\omega^3}h(x,y),\label{37}\en
where we have introduced, to simplify the notation, $x=S^0$ and
$y=S^+S^-$. In particular, the existence of a superconducting
phase corresponds to the existence of a non trivial zero of $f_1$
and $f_2$, \cite{bm,martin}. Due to the definition of $f_1$ and
$f_2$ it is clear that a pair $(x_o,y_o)$, with $x_o\neq 0$ and
$y_o\neq 0$, is such that $f_1(x_o,y_o)=f_2(x_o,y_o)=0$ if and
only if it is a zero of the function $h$: $h(x_o,y_o)=0$. In order
to find such a zero, it is first necessary to obtain an explicit
expression for the coefficients $\Re\Gamma_\pm^{(\gamma)}$. This
is easily done using the definitions in (\ref{223}), since we get
\be \Re\Gamma_\pm^{(a)}=\frac{1}{2}\int_{-\infty}^\infty\sum_{\vec
p\in\Lambda_N}|f_m(\vec p)|^2e^{-i\tau\nu_\pm(\vec
p)}\,d\tau=\pi\sum_{\vec p\in\Lambda_N}|f_m(\vec
p)|^2\delta(\nu_\pm(\vec p)),\label{38}\en and \be
\Re\Gamma_\pm^{(b)}=\pi\sum_{\vec p\in\Lambda_N}|f_n(\vec
p)|^2\delta(\nu_\pm(\vec p)).\label{39}\en It is now almost
straightforward to recover the results of \cite{bm,martin}.
Following Buffet and Martin's original idea, we look for solutions
corresponding to $\nu=0$. This means that, because of (\ref{212}),
the value of $x=S^0$ is fixed: $x=-2\tilde\epsilon/g$. Moreover,
with this choice, $\nu_+(\vec p)=\omega-\epsilon_{\vec p}$, which
is zero if and only if $\omega=\epsilon_{\vec p}$. Also, we have
$\nu_-(\vec p)=-\omega-\epsilon_{\vec p}$, which is never zero.
For these reasons we deduce that $\Re\Gamma_-^{(\gamma)}=0$,
$\gamma=a,b$, while the sums in (\ref{38}) and (\ref{39}) for
$\Re\Gamma_+^{(\gamma)}$ are restricted to the smaller set,
$\E_N\subset\Lambda_N$, of those values of $\vec p$ such that, if
$\vec q\in\E_N$ then $\epsilon_{\vec q}=\omega$. Therefore,
recalling the expression of $m(\vec p)$ and $n(\vec p)$ in
(\ref{215bis}), we find \be
\Re\Gamma_+^{(a)}=\pi\frac{e^{\beta\omega}}{e^{\beta\omega}-1}\sum_{\vec
p\in\E_N}|f(\vec p)|^2, \hspace{4mm}
\Re\Gamma_+^{(b)}=\pi\frac{1}{e^{\beta\omega}-1}\sum_{\vec
p\in\E_N}|f(\vec p)|^2, \label{310}\en

so that equation $f(x,y)=0$ looks like
$$ \pi\frac{e^{\beta\omega}}{e^{\beta\omega}-1}\sum_{\vec
p\in\E_N}|f(\vec p)|^2\frac{\omega-g}{(\omega+gx)^2}+
\pi\frac{1}{e^{\beta\omega}-1}\sum_{\vec p\in\E_N}|f(\vec
p)|^2\frac{\omega+g}{(\omega+gx)^2}=0,
$$
or \be e^{\beta\omega}=\frac{g+\omega}{g-\omega}.\label{311}\en
This equation is the crucial one, which replaces the one obtained
in \cite{bm,martin}, $g\tanh\left(
\frac{\beta\omega}{2}\right)=\omega$. As a matter of fact, in
\cite{bagbcs} it is also proven that these two equations are
equivalent, and for that we recover exactly the same values of the
critical temperature and of the order parameters as in \cite{bm}.

\subsection{More results}

We have shown how the SLA can be successfully used to analyze the
phase structure of low temperature superconductivity analyzing a
strong coupling BCS model, considered as an open system
interacting with a bosonic thermal bath.

This procedure is rather direct and  technically much simpler than
the one used in  \cite{bm}. Among the other simplifications, for
instance, a single equation $h(x,y)=0$ must be solved instead of
the system $f_1(x,y)=f_2(x,y)=0$. This suggests to use the SLA
also to modify the original model in the attempt of getting some
insight on high-temperature superconductivity. This project
started quite recently, \cite{bagbcs2}, by introducing two
reservoirs instead of only one, as we did here, to see whether the
value of the critical temperature increases because of the
presence of this second reservoir. Our results seem rather
promising but not jet definitive. A deeper analysis is being
presently undertaken.

\vspace{6mm}

\noindent{\large \bf Acknowledgments} \vspace{3mm}

I  would like to acknowledge financial support by the Murst,
within the  project {\em Problemi Matematici Non Lineari di
Propagazione e Stabilit\`a nei Modelli del Continuo}, coordinated
by Prof. T. Ruggeri.

\newpage

\appendix
\renewcommand{\theequation}{\Alph{section}.\arabic{equation}}


 \section{\hspace{-14.5mm} Appendix:  Few results on the stochastic limit}

In this Appendix we will briefly summarize some of the basic facts
and properties concerning the SLA which are used all throughout
the paper. We refer to \cite{book} and references therein for more
details.

Given an open system ${\cal S}+{\cal R}$ we write its hamiltonian
$H$ as the sum of two contributions, the free part $H_0$ and the
interaction $\lambda H_I$. Here $\lambda$ is a coupling constant,
$H_0$ contains the free evolution of both the system ${\cal S}$
and the reservoir ${\cal R}$, while $H_I$ contains the interaction
between ${\cal S}$ and ${\cal R}$. Working in the interaction
picture, we define $H_I(t)=e^{iH_0t}H_Ie^{-iH_0t}$ and the so
called wave operator $U_\lambda(t)$ which is the solution of the
following differential equation \be
\partial_t U_\lambda(t)=-i\lambda H_I(t)U_\lambda(t),
\label{a1} \en with the initial condition $U_\lambda(0)=\Id$.
Using the van-Hove rescaling $t\rightarrow \frac{t}{\lambda^2}$,
see \cite{martin,book} for instance, we can rewrite the same
equation in a form which is more convenient for our perturbative
approach, that is \be
\partial_t U_\lambda(\frac{t}{\lambda^2})=-\frac{i}{\lambda} H_I(\frac{t}{\lambda^2})U_\lambda(\frac{t}{\lambda^2}),
\label{a2} \en with the same initial condition as before. Its
integral counterpart is \be
U_\lambda(\frac{t}{\lambda^2})=\Id-\frac{i}{\lambda} \int_0^t
H_I(\frac{t'}{\lambda^2})U_\lambda(\frac{t'}{\lambda^2})dt',
\label{a3} \en which is the starting point for a perturbative
expansion, which works in the following way.

Suppose, to begin with, that we are interested to the zero
temperature situation. Then let $\varphi_0$ be the ground vector
of the reservoir and $\xi$ a generic vector of the system. Now we
put $\varphi_0^{(\xi)}=\varphi_0\otimes\xi$. We want to compute
the limit, for $\lambda$ going to $0$, of the first non trivial
order of the mean value of the perturbative expansion of
$U_\lambda(t/\lambda^2)$ above in $\varphi_0^{(\xi)}$, that is the
limit of \be I_\lambda(t)=(-\frac{i}{\lambda})^2\int_0^t dt_1
\int_0^{t_1}dt_2\langle
H_I(\frac{t_1}{\lambda^2})H_I(\frac{t_2}{\lambda^2})\rangle_{\varphi_0^{(\xi)}},
\label{a4} \en for $\lambda\rightarrow 0$. Under some regularity
conditions on the functions which are used to smear out the
(typically) bosonic fields of the reservoir, this limit is shown
to exist for many relevant physical models, see \cite{book}, and
\cite{bagacc,baglaser,bagbcs} for few recent applications to
quantum many body theory. It is at this stage that all the complex
quantities like the  $\Gamma_\alpha^{(\gamma)}$'s we have
introduced in the main body of this paper appear. We define
$I(t)=\lim_{\lambda\rightarrow 0}I_\lambda(t)$. In the same sense
of the convergence of the (rescaled) wave operator
$U_\lambda(\frac{t}{\lambda^2})$ (the convergence in the sense of
correlators), it is possible to check that also the (rescaled)
reservoir operators converge and define new operators which do not
satisfy canonical commutation relations but a modified version of
these. For instance, in Section II this procedure has produced the
operators $b_{\alpha,\beta}(t)$ starting from $b(\vec k)$.
Moreover, these limiting operators depend explicitly on time and
they live in a Hilbert space which is different from the original
one. In particular, they annihilate a vacuum vector, $\eta_0$,
 which is no longer the original one, $\varphi_0$. This is what happens, for instance, if $\varphi_0$  depends on
 $\lambda$, $\varphi_0\rightarrow \varphi_0^{(\lambda)}$, and
 considering $\eta_0$ as the following limit: $\eta_0=\lim_{\lambda\rightarrow
 0}\varphi_0^{(\lambda)}$.

It is not difficult to deduce the form of a time dependent
self-adjoint operator $H_I^{(sl)}(t)$, which depends on the system
operators and on the limiting operators of the reservoir, such
that the  first non trivial order of the mean value of the
expansion of $U_t=\Id-i\int_0^tH_I^{(sl)}(t')U_{t'}dt'$ on the
state $\eta_0^{(\xi)}=\eta_0\otimes\xi$ coincides with $I(t)$. The
operator $U_t$  defined by this integral equation is called again
the {\em wave operator}.

The form of the generator follows now from an operation of normal
ordering. More in details, we start defining the flux of an
observable  $\tilde X=X\otimes \Id_{r}$, where $\Id_{r}$ is the
identity of the reservoir and $X$ is an observable of the system,
as $j_t(\tilde X)=U_t^\dagger \tilde XU_t$. Then, using the
equation of motion for $U_t$ and $U_t^\dagger$, we find that
$\partial_t j_t(\tilde X)=iU_t^\dagger [H_I^{(sl)}(t),\tilde
X]U_t$. In order to compute the mean value of this equation on the
state $\eta_0^{(\xi)}$, so to get rid of the reservoir operators,
it is convenient to compute first the commutation relations
between $U_t$ and the limiting operators of the reservoir. At this
stage the so called {\em time consecutive principle} is used in a
very heavy way to simplify the computation. This principle, which
has been checked for many classes of physical models, \cite{book},
states that, if $\beta(t)$ is any of these limiting operators of
the reservoir, then \be [\beta(t),U_{t'}]=0, \mbox{ for all }
t>t'. \label{a5} \en Using this principle and recalling that
$\eta_0$ is annihilated by the limiting annihilation operators of
the reservoir, it is now a simple exercise to compute
$\langle\partial_t j_t(X)\rangle_{\eta_0^{(\xi)}}$ and, by means
of the equation $\langle\partial_t
j_t(X)\rangle_{\eta_0^{(\xi)}}=\langle
j_t(L(X))\rangle_{\eta_0^{(\xi)}}$, to identify the form of the
generator of the physical system.

\vspace{4mm}

Let us now consider the case in which $T>0$. In this case the
state of the reservoir is no longer given by the vacuum
$\varphi_0$. It is now convenient to use the so-called {\em
canonical representation of thermal states}, \cite{book}. Using
the same notation of Section IV, any annihilator operator $a_{\vec
p,j}$ can be written as the following linear combination \be
a_{\vec p,j}=\sqrt{m(\vec p)}\,c_{\vec p,j}^{(a)}+\sqrt{n(\vec
p)}\,c_{\vec p,j}^{(b),\dagger},\label{a6}\en where $m(\vec p)$
and $n(\vec p)$ are the following two-points functions, \be m(\vec
p)=\omega_\beta(a_{\vec p,j}a_{\vec
p,j}^\dagger)=\frac{1}{1-e^{-\beta\epsilon_{\vec p}}},
\hspace{1cm}n(\vec p)=\omega_\beta(a_{\vec p,j}^\dagger a_{\vec
p,j})=\frac{e^{-\beta\epsilon_{\vec p}}}{1-e^{-\beta\epsilon_{\vec
p}}},\label{a6bis}\en for our bosonic reservoir, if $\omega_\beta$
is a KMS state corresponding to an inverse temperature $\beta$.
The operators $c_{\vec p,j}^{(\alpha)}$ are assumed to satisfy the
following commutation rules \be [c_{\vec p,j}^{(\alpha)},{c_{\vec
q,k}^{(\gamma)}}^\dagger]=\delta_{jk}\delta_{\vec p\,\vec
q}\delta_{\alpha\gamma}, \label{a7}\en while all the other
commutators are trivial. Let moreover $\Phi_0$ be the vacuum of
the operators $c_{\vec p,j}^{(\alpha)}$:
$$
c_{\vec p,j}^{(\alpha)}\Phi_0=0, \hspace{1cm}\forall \vec
p,j,\alpha.
$$
Then it is immediate to check that the results in (\ref{a6bis})
for the KMS state can be found, using these new variables,
representing $\omega_\beta$ as the following  vector state
$\omega_\beta(\cdot)=\langle\Phi_0,\cdot\Phi_0\rangle$. With this
GNS-like representation it is trivial to check that both the CCR
and the two-point functions are easily recovered. This
representation is also called in \cite{book} the Fock-anti Fock
representation because of the different sign in the free time
evolution of the annihilation operators $c_{\vec p,j}^{(a)}$ and
$c_{\vec p,j}^{(b)}$. Once this representation is introduced, all
the same steps as for the situation with $T=0$  can  be repeated,
and the expression for the generator can be deduced using exactly
the same strategy.

\end{document}